\documentclass[twocolumn,showpacs,prb]{revtex4}
\usepackage{graphicx}

\begin{document}
\title{Spin Transitions Induced by Magnetic Field in  
Quantum Dot Molecules}

\author{Ramin M. Abolfath$^{1,2}$, and Pawel Hawrylak$^1$}

\affiliation{
$^1$ Institute for Microstructural Sciences,
National Research Council of Canada,
Ottawa, K1A 0R6, Canada \\
$^2$ Department of Radiation Oncology, University of Texas Southwestern
Medical Center, Dallas, Texas 75390, USA
}

\date{\today}

\begin{abstract}
We present a theoretical study of 
magnetic field driven spin transitions of electrons in  
coupled lateral quantum dot molecules.  
A detailed numerical study of spin phases of artificial molecules 
composed of two laterally coupled quantum dots with
$N=8$ electrons  is presented as a function of 
magnetic field, 
Zeeman energy, and the detuning  using real space Hartree-Fock 
Configuration Interaction (HF-CI) technique. 
A microscopic picture of quantum Hall ferromagnetic phases 
corresponding to zero and full spin polarization at filling 
factors $\nu=2$ and $\nu=1$,
and ferrimagnetic phases resulting from coupling of the two dots,
is presented. 

\end{abstract}

\maketitle

\section{Introduction}
The application of spin of electrons in quantum dots for generation of  
electron entanglement and quantum information processing 
in solid state devices is of current experimental 
\cite{kouwenhoven,Ciorga,MichelPRL93,petta,Korkusinski,petta2,Koppens,
GuyPhysicaB,Kouwenhoven2,Hanson03,GuyPRB} 
and theoretical interest
\cite{JJP,qinfo,brum,Loss-DiVincenzo,HuDasSarma,Kolehmainen,RaminWojtekPawel}.
Controlling the spin of electrons in single quantum dots by tuning the
external magnetic field, the confining potential, 
number of electrons, and Zeeman coupling has been demonstrated 
\cite{Ciorga,MichelPRL93,petta,Korkusinski,petta2,Koppens}.
It was shown that in strong magnetic field  electrons form 
a spin singlet quantum Hall droplet at filling factor $\nu=2$.
Increasing the magnetic field leads to the spin-flip transitions until
the spin polarized filling factor $\nu=1$ droplet is reached.\cite{Ciorga}
Spin flips beyond the first spin flip are associated with  
correlated states such as spin bi-excitons, 
identified and observed experimentally  
\cite{Korkusinski}.
Quantum dot molecules offer additional possibility of coupling and 
controlling spin transitions by tuning the 
tunneling barrier which controls the inter-dot coupling. 
\cite{HuDasSarma,Kolehmainen,RaminWojtekPawel,JJP,Canted}
The recently demonstrated time dependent control of the tunneling 
barrier height and confining potential \cite{petta2}, and 
the quantum state of the electron spin by applying oscillating
magnetic field (Rabi oscillations) \cite{Koppens}, resulted in 
coherent manipulation of two electron spins in coupled quantum dot
molecules.
Recent experiments by Pioro-Ladriere {\em et al.} in 
Ref.\onlinecite{MichelPRL93}
suggested that in strong magnetic field electrons are expected to form 
quantum Hall droplet in each quantum dot. 
Edge states of each droplet can be coupled  
in a controlled way using barrier electrodes, and at filling factor $\nu=2$  
effectively reduce the many-electron-double dot system to a 
two-level molecule \cite{MichelPRL93}. 
When populated with one electron each, 
one expects to have singlet-triplet transitions of two valence electrons in the
background of core electrons of the spin singlet $\nu=2$ droplets. 
With increasing magnetic field, transitions to higher spin polarized states
are expected, with coupled quantum dots 
resembling artificial magnetic molecules.

In this paper, we investigate the effect of the interdot tunneling
and electron-electron interactions on the evolution of total spin of 
electrons in a quantum dot molecule as a function of electron numbers 
and magnetic field.
We study the many-body effects in the spin flip transitions
by incorporating systematically the inter-dot and intra-dot 
electron-electron Coulomb interactions using real space Hartree-Fock 
Configuration Interaction (HF-CI) technique.
We find quantum Hall droplets  with zero and full spin polarization, 
identified as $\nu=2$ and
$\nu=1$ quantum Hall droplets \cite{JJP},
in analogy with single quantum dots
and quantum Hall ferromagnets \cite{SteveAllan}.
Between these two states, we find 
series of continuous transitions among partially spin polarized phases.
These partially polarized phases correspond to spin flips. Simultaneous 
spin flip in each isolated dot must lead to even number of spin flips 
in a double dot. 
Recently, we have found partially spin polarized phases
which correspond to odd number of spin flips\cite{QHFerriPRL} in a double quantum dot.
In Ref. [\onlinecite{QHFerriPRL}],
we have identified these correlated states as quantum Hall {\em ferrimagnets}.

Coherent superposition of two single particle levels in a double  well potential
in a form of symmetrical and antisymmetrical states is a well known
example of  quantum mechanical phenomena, from
coherent charge oscillation of an electron between two  states  localized
on two protons in 
$H^+_2$ molecules,  coherent oscillation between left handed
and right handed Amino acids, \cite{Sakurai}
coherent control of Rabi oscillations of electron spin 
in quantum dots \cite{Koppens}, to
 macroscopic quantum resonance of
 Cooper pairs  in mesoscopic
superconducting grains \cite{SC grains}.
The quantum Hall ferrimagnetic states, or spin unbalanced phases,
\cite{Ferri1,Ferri2}
are also a direct manifestation of coherent
quantum mechanical tunneling and inter-dot electronic correlations.
These states can be described in terms of linear combinations
of spin excitons localized in left and right dots, which in turn
lead to coherent spin oscillations, e.g.,
spin counter-part of coherent charge
oscillations in $H^+_2$ molecules.
 
The paper is organized as follows:
In section \ref{quantumdots} we review the Hamiltonian 
of electrons confined in the lateral gated quantum dots.
In section \ref{SP-HFCI},
the computational methods of single particle configuration interaction
(SP-CI) and unrestricted Hartree-Fock configuration interaction (URHF-CI)
are summarized.
To differentiate the spin transitions of quantum dot 
molecules and two isolated dots with zero inter-dot interaction,
we briefly present the spin phase diagram of single 
dots in section \ref{SDsec}.
The microscopic picture of spin excitations in 
coupled quantum dots are discussed in sections \ref{SpinTransitions},
and \ref{realspace}.
The interpretation of spin excitations in terms of electron-hole 
excitations allows us to attribute the excitations with total spin $S=1$ and $S=2$ 
as spin exciton and bi-exciton.
Pairing of  excitons and the formation of
bi-excitons due to strong inter-dot interaction is discussed.
To elucidate the quantum Hall ferrimagnetic states,   
the real space representation of excitons is introduced.
%
%
The paper is summarized in section \ref{Conclusion}.

\section{Hamiltonian}
\label{quantumdots}
We describe electrons confined in quasi-two-dimensional 
quantum dots in a uniform perpendicular 
magnetic field by the effective mass Hamiltonian
\begin{eqnarray}
H = \sum_{i=1}^N  \left( T_i + E_{iZ} \right)
+ \frac{e^2}{2\epsilon}\sum_{i \neq j}\frac{1}{|\vec{r}_i - \vec{r}_j|},
\label{ham1}
\end{eqnarray}
where
\begin{eqnarray}
T=\frac{1}{2m^*}\left(\frac{\hbar}{i}\vec{\nabla}
+ \frac{e}{c} A(\vec{r})\right)^2 + V(x,y)
\label{ham2}
\end{eqnarray}
is the single electron Hamiltonian in magnetic field. Here $(\vec{r})=(x,y)$
describes electron position, $A(\vec{r})=\frac{1}{2}\vec{B}\times\vec{r}$ 
is the vector potential, $B$ is the external magnetic field, and
$V(\vec{r})$ is the quantum dots confining potential.
$m^*$ is the conduction-electron effective mass, 
$e$ is the electron charge, and
$\epsilon$ is the host semiconductor dielectric constant 
($\epsilon=12.8$ in GaAs).
$E_{iZ}= \frac{1}{2}g\mu_B \sigma_{iz} B$ is the Zeeman spin splitting, 
$g$ is the host semiconductor $g$-factor ($g=-0.44$ in GaAs),
$\mu_B$ is the Bohr magneton, and $\sigma$ is the Pauli matrix.
In what follows,
we present the numerical results in effective atomic unit
(in GaAs effective Bohr radii $a^*_0= 9.79 nm$, 
and effective Rydberg $Ry^*=5.93 meV$).

The single particle eigenvalues ($\epsilon_i$) and 
eigenvectors ($\varphi_i$) are calculated by
discretizing $T$ in real space, and diagonalizing the resulting matrix
using conjugate gradient algorithms \cite{RaminWojtekPawel}.
The details of this calculation can be found in Ref. [\onlinecite{JCP06}].

\section{Many body spectrum}\label{SP-HFCI}
To calculate the spectrum of interacting electrons, described
by Hamiltonian $H$ in Eq.(\ref{ham1}), we employ either the real space
single particle or unrestricted Hartree-Fock states in 
configuration interaction techniques.\cite{JCP06}
In the first SP-CI approach, single-particle levels are used to 
construct many-electron configurations which are the basis of 
configuration interaction (CI) Hamiltonian.
Denoting the creation (annihilation) operators for electrons in
non-interacting SP state $|\alpha\sigma\rangle$ by 
$c^\dagger_{\alpha\sigma}~(c_{\alpha\sigma})$,
the Hamiltonian of an interacting electron system in second quantization
can be written as
\begin{eqnarray}
H&=&\sum_{\alpha}\sum_{\sigma} 
\epsilon_{\alpha} 
c^\dagger_{\alpha\sigma} c_{\alpha\sigma} 
\nonumber \\ &&
+ \frac{1}{2}\sum_{\alpha\beta\gamma\mu}\sum_{\sigma\sigma'}
V_{\alpha\sigma,\beta\sigma',\gamma\sigma',\mu\sigma}
c^\dagger_{\alpha\sigma} c^\dagger_{\beta\sigma'} 
c_{\gamma\sigma'} c_{\mu\sigma} 
\label{multiparticle0}
\end{eqnarray}       
where the first term is the single particle Hamiltonian,
and
$V_{\alpha\sigma,\beta\sigma',\gamma\sigma',\mu\sigma} = 
\int d\vec{r} \int d\vec{r'}  
\varphi^*_{\alpha\sigma}(\vec{r})
\varphi^*_{\beta\sigma'}(\vec{r'}) 
\frac{e^2}{\epsilon|\vec{r}-\vec{r'}|}
\varphi_{\gamma\sigma'}(\vec{r'})
\varphi_{\mu\sigma}(\vec{r})$, 
is the two-body Coulomb matrix element.
In the configuration interaction method 
the Hamiltonian of an 
interacting system is calculated in the basis of finite number of
many-electron configurations. The total number of configurations 
(or Slater determinants participating in CI calculation)
is determined by
\begin{eqnarray}
N_C = \left[\frac{N_s!}{N_\uparrow!(N_s-N_\uparrow)!}\right]
\left[\frac{N_s!}{N_\downarrow!(N_s-N_\downarrow)!}\right].
\end{eqnarray}
Here $N_s$ is the number of single particle levels, and
$N_\uparrow$ and $N_\downarrow$ are the number of spin up 
and spin down electrons.
This Hamiltonian is either diagonalized exactly for small systems
or low energy eigenvalues and eigenstates are extracted 
approximately for very large number of configurations. \cite{JCP06}
With increasing the number of single-particle levels $N_s$, 
the number of configurations $N_C$ grows very fast, yet a large
number is needed to accurately account for direct and exchange interaction, 
and electronic correlations.
To improve the convergence of CI method we incorporate direct and exchange
contribution into the basis states
by replacing SP states with states obtained by the 
unrestricted Hartree-Fock method (URHF-CI).
This implies expressing the new creation (annihilation) 
operators for URHF quasi-particles by $a^\dagger_{i\sigma}$ 
($a_{i\sigma}$), with
the index $i$ representing the URHF orbit quantum numbers.
The URHF basis can be expanded in a linear combination of SP states.
In terms of SP creation (annihilation) operators we write  
\begin{eqnarray}
a^\dagger_{i\sigma}= \sum_{\alpha=1}^{N_l} 
\lambda^{(i)}_{\alpha\sigma} c^\dagger_{\alpha\sigma}.
\end{eqnarray}
The transformation coefficients, $\lambda^{(i)}_{\alpha\sigma}$, satisfy 
the self-consistent Pople-Nesbet equations 
\cite{Szabo_book,YannouleasPRB,RaminWojtekPawel,JCP06}:
\begin{eqnarray}
&&\sum_{\gamma=1}^{N_l}
\{\epsilon_\mu\delta_{\gamma\mu}+
\sum_{\alpha,\beta=1}^{N_l} (V_{\mu\alpha\beta\gamma}
\sum_{\sigma'}
\sum_{j=1}^{N_{\sigma'}}
\lambda^{*(j)}_{\alpha\sigma'} \lambda_{\beta\sigma'}^{(j)} 
\nonumber \\&&
-V_{\mu\alpha\gamma\beta}
\sum_{\sigma'}\sum_{j=1}^{N_{\sigma'}}
\lambda^{*(j)}_{\alpha\sigma'} \lambda_{\beta\sigma'}^{(j)}
\delta_{\sigma,\sigma'} ) \} 
\lambda_{\gamma\sigma}^{(i)} 
= \epsilon^{HF}_{i\sigma} ~ \lambda_{\mu\sigma}^{(i)},
\label{urhfeq1}
\end{eqnarray}
where $\epsilon^{HF}_{i\sigma}$ are the URHF eigenenergies.
The $N$-lowest energy URHF levels form a Slater determinant occupied
by HF quasi-electrons, corresponding to HF ground state.
The rest of orbitals with higher energies are outside of the HF
Slater determinant (unoccupied HF levels), contribute to electronic
correlations, and can be used for CI calculation.
The many body Hamiltonian of the interacting system 
in the URHF basis can finally be written as:
\begin{eqnarray}
H&=&\sum_\sigma\sum_{ij} 
\langle i\sigma | T | j\sigma \rangle 
a^\dagger_{i\sigma} a_{j\sigma} \nonumber \\&&
+ \frac{1}{2}\sum_{ijkl}\sum_\sigma\sum_{\sigma'}
U_{i\sigma,j\sigma',k\sigma',l\sigma}
a^\dagger_{i\sigma} a^\dagger_{j\sigma'} 
a_{k\sigma'} a_{l\sigma} ,
\label{multiparticleCI}
\end{eqnarray}       
where 
$U_{i\sigma,j\sigma',k\sigma',l\sigma}$ 
are the Coulomb matrix elements in the URHF basis.
Here 
\begin{equation}
\langle i\sigma | T | j\sigma \rangle = \epsilon^{HF}_{i\sigma} \delta_{ij} -
\langle i\sigma | U_H - U_X | j\sigma \rangle,
\end{equation}       
where $U_H$ and $U_X$ are the Hartree and exchange operators
\begin{eqnarray}
\langle i \sigma | U_H  | j \sigma \rangle = 
\sum_{\alpha\beta\mu\nu=1}^{N_l} V_{\alpha\mu\gamma\beta}
\lambda_{\mu\sigma}^{*(i)} \lambda_{\gamma\sigma}^{(j)}
\sum_{\sigma'} \sum_{k=1}^{N_{\sigma'}}
\lambda_{\alpha\sigma'}^{*(k)} \lambda_{\beta\sigma'}^{(k)},\nonumber
\end{eqnarray}
\begin{eqnarray}
\langle i\sigma | U_X  | j\sigma \rangle = 
\sum_{\alpha\beta\mu\nu=1}^{N_l} V_{\alpha\mu\beta\gamma}
\lambda_{\alpha\sigma}^{*(i)} \lambda_{\beta\sigma}^{(j)}
\sum_{k=1}^{N_{\sigma}}
\lambda_{\mu\sigma}^{*(k)} \lambda_{\gamma\sigma}^{(k)}.\nonumber
\end{eqnarray}
The resulting CI Hamiltonian matrix constructed in the basis of 
URHF configurations  is either diagonalized exactly for small systems,
or low energy eigenvalues and eigenstates are extracted approximately 
for very large number of configurations \cite{JCP06}.
The details of the calculation and the convergence of the results as
a function of number of basis and configurations, and comparison between
SP-CI and URHF-CI methods
can be found in Ref. \onlinecite{JCP06}.

\section{spin transitions in a single dot}\label{SDsec}
\begin{figure}
\begin{center}
\includegraphics[width=0.8\linewidth]{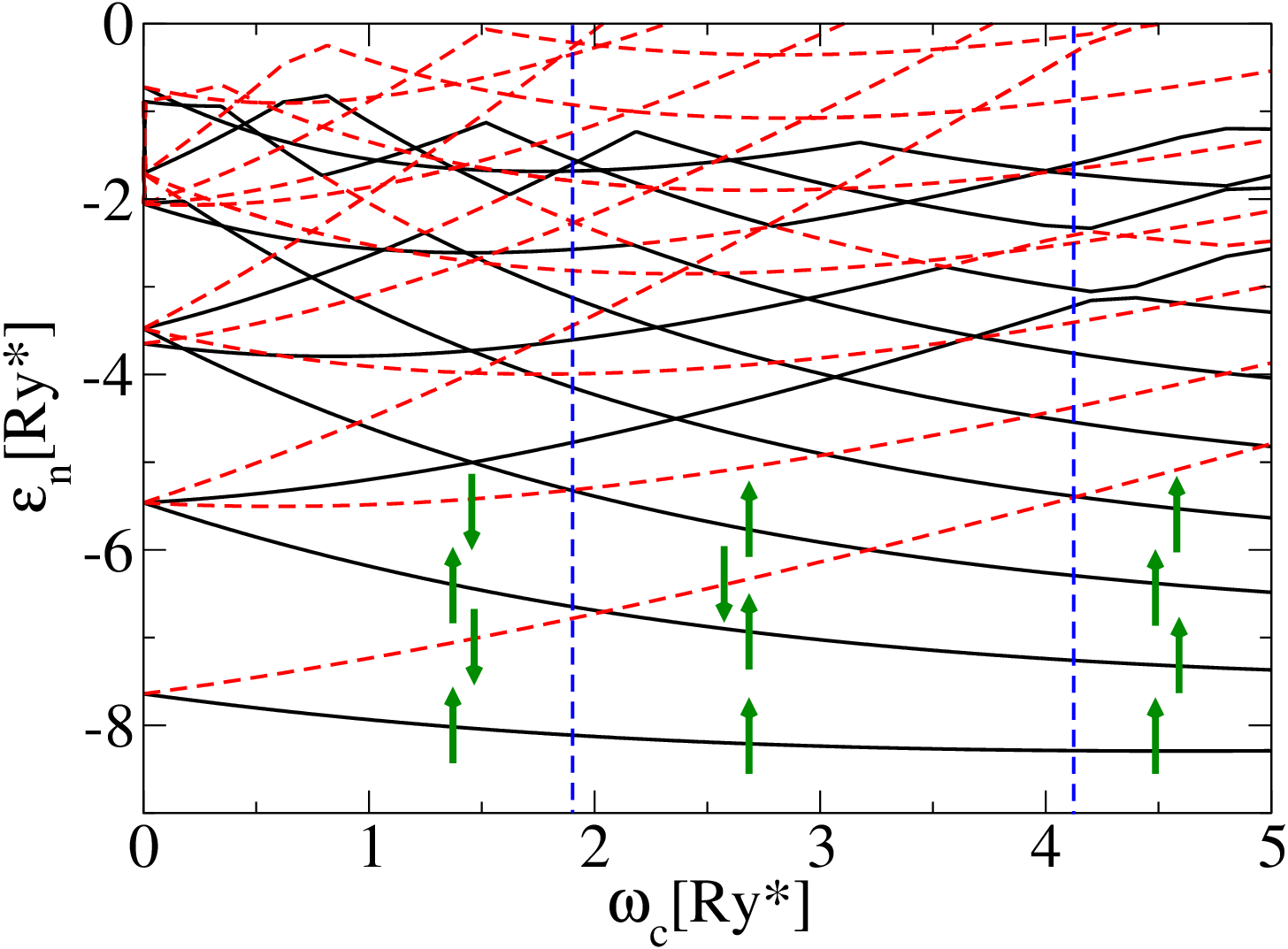}
\noindent
\caption{
(Color online) Single particle spectrum as a function of 
cyclotron frequency $\omega_c$
for a gaussian single dot with strength $V_0=-10Ry^*$, and
$\Delta=2.5 a_0^*$ in the presence of Zeeman splitting. 
Arrows represent spin of electrons.
For illustration purposes a very high Zeeman coupling $g=-9$ is used.
}
\label{spectrum_SD}
\end{center}
\end{figure}

We describe a single dot by an isotropic gaussian confining potential 
$V(x,y) = V_0 e^{-(x^2 + y^2)/\Delta^2}$.
The single particle eigenenergies of such quantum dot as a function of
cyclotron frequency $\omega_c=eB/m^*$ are shown in Fig. \ref{spectrum_SD}.

States with spin up (down) are shown by bold (dashed) lines.
For illustration, a very large g-factor is introduced. 
We note in Fig. \ref{spectrum_SD}, with increasing magnetic field
the energy of spin up (bold) levels decrease. 
These levels, and their
spin down partners, correspond to the levels of the lowest Landau level (LLL). 
We now populate the lowest energy states with a number of electrons. 
From previous work \cite{Korkusinski}, the minimum number of electrons which 
exhibits all nontrivial 
phenomena in the spin evolution of a single quantum dot is $N=4$.
The $N=4$ configurations which minimize the
kinetic and Zeeman energy are shown in Fig.\ref{spectrum_SD}. Due to the crossing
of spin up and down levels, there are three different configurations 
 $S_z=0$, $S_z=1$, and $S_z=2$. These configurations illustrate increasing 
spin polarization of the electronic droplet with increasing magnetic field.
With very small Zeeman energy the increasing spin polarization 
in quantum dots  is driven by electron-electron interactions.
We hence turn off the 
Zeeman coupling and  turn on electron-electron interactions.

\begin{figure}
\begin{center}
\includegraphics[width=0.8\linewidth]{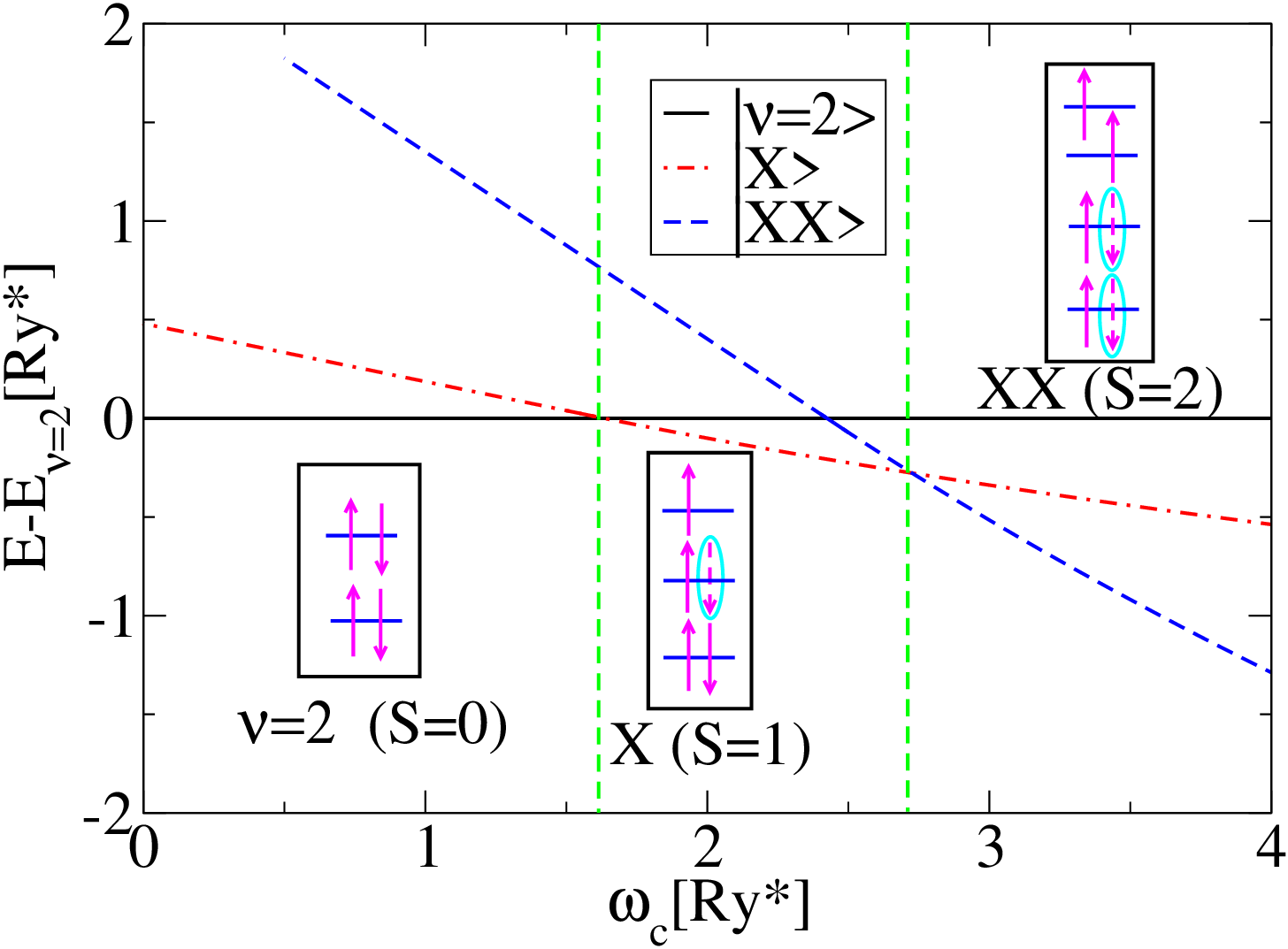}
\noindent
\caption{
(Color online) The energy of spin configurations shown in the boxes, 
$|\nu=2\rangle$, $|X\rangle$, and $|XX\rangle$ with total spin 
$S=0$, $S=1$, and $S=2$,
using LLL orbitals, and $E_Z=0$.
$E_{\nu=2}$ ($S=0$) is the reference of energy.
The arrows surrounded by circles represent the holes.
}
\label{SD_SP_CI}
\end{center}
\end{figure}

We start with the lowest energy configuration build with SP LLL states,
the $S_z=0$ spin singlet $\nu=2$ configuration $|\nu=2\rangle$.
The spin excitations with $S=1$ and $S=2$, are constructed by removing
electrons from occupied states and putting to unoccupied states, and
can be described in terms of
single exciton $|X\rangle$ and biexciton $|XX\rangle$ configurations,
as  shown in Fig. \ref{SD_SP_CI}.
Neglecting the mixing between configurations, we calculate the energy
of each spin configuration.
The result of this calculation is shown in Fig. \ref{SD_SP_CI}. 
We chose the energy of the $S=0,\nu=2$ state ($E_{\nu=2}$) as 
the reference energy.
As shown in Fig. \ref{SD_SP_CI},
with increasing magnetic field even without Zeeman energy
both the first and second spin flip transitions 
occur at $\omega_c=1.6$ and $\omega_c=2.75$ 
due to electron-electron interactions.

\begin{figure}
\begin{center}
\includegraphics[width=0.8\linewidth]{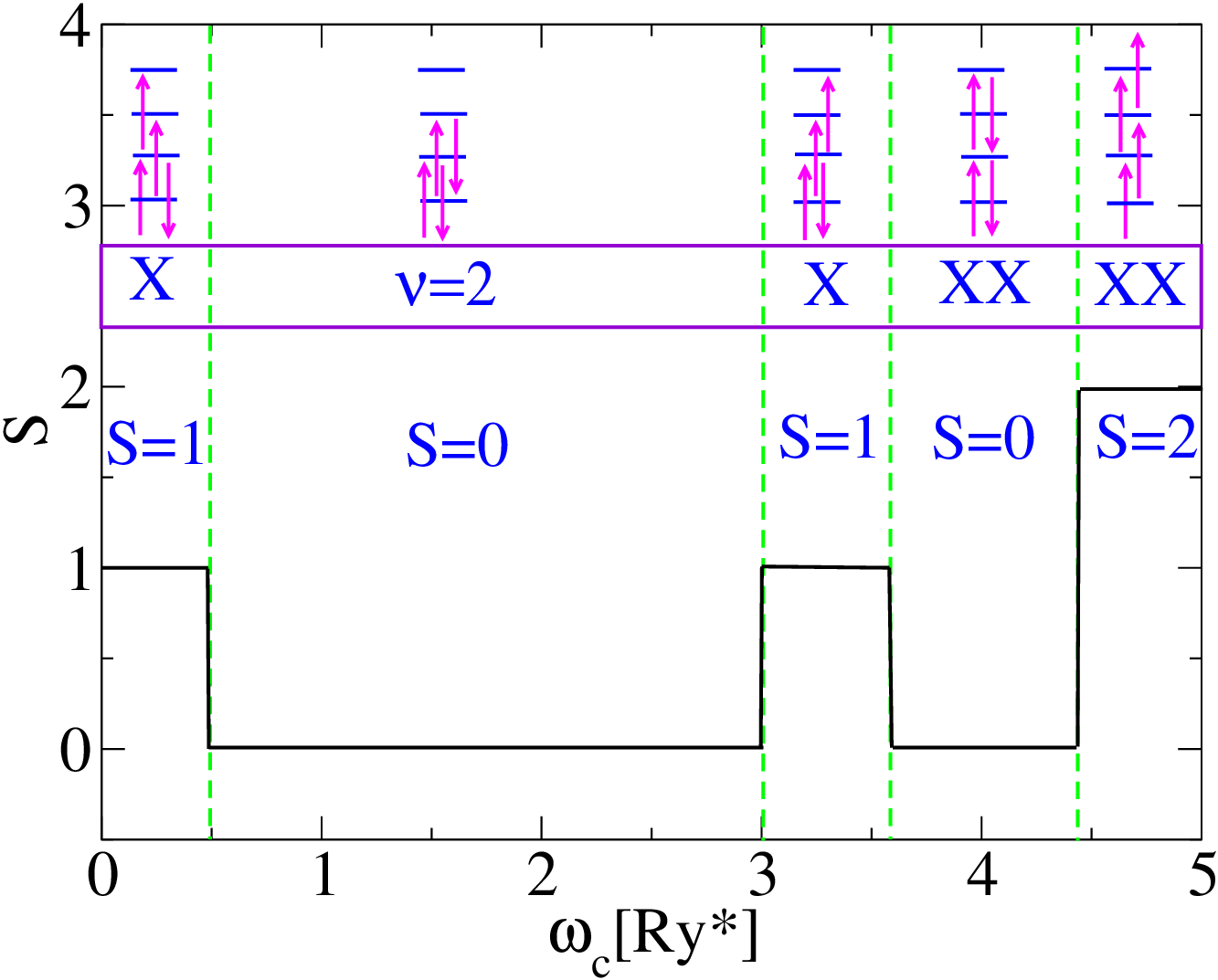}
\noindent
\caption{
(Color online) The spin evolution of the ground state of
single quantum dot as a function of magnetic field 
using URHF-CI method with $N_s=8$ HF levels, corresponding to 
$N_C=784$ configurations.
}
\label{SD_HFCI}
\end{center}\vspace{-0.5cm}
\end{figure}

The effect of correlations on the evolution of 
spin of electrons in a single quantum dot obtained by the URHF-CI
method is shown in Fig. \ref{SD_HFCI}.
Here the spin unpolarized ($S_z=0$) URHF states 
have been constructed out of $N_l=30$ SP states.
$N_s=8$ HF levels have be taken to construct CI Hamiltonian,
resulting in $N_C=784$ configurations.
At zero magnetic field we find $S=1$ triplet
due to Hund's rule for electrons in a half-filled p-shell.
With increasing magnetic field, the single particle energy gap opens up,
leading to suppression of the $S=1$ triplet state, and formation of the 
$\nu=2$ $(S=0)$ singlet state.
The first spin flip $S=1$  state appears around
$\omega_c=3$, followed by  the second spin flip spin polarized  $S=2$ 
state at $\omega_c=4.5$. 
The flipping of the second spin is 
interrupted by a low-spin, $S=0$, strongly correlated state. This state was 
previously identified with the formation of spin singlet 
bi-exciton.\cite{Korkusinski} 
The first and second spin flip state can be 
obtained both for  the noninteracting electrons and in
Hartree-Fock  approximation while the formation of spin singlet 
bi-exciton is a result of electronic correlations in a quantum dot.

\section{spin transitions in quantum dot molecules}
\label{SpinTransitions}

We now turn to study the spin transitions  
of laterally coupled quantum dots. We describe the molecule by
electron $(N_L,N_R)$ and ground state spin numbers $(S_L,S_R)$
of isolated left (L) and right (R) dots. 
The spin phase diagram turns out
to depend on electron numbers in each dot. Here we will focus on 
molecules build out of identical dots with $N_L=N_R$. For a given number of 
electrons the magnetic field will be used to tune their individual 
spin $S_L=S_R$. 
The goal will be to determine the total spin of the molecular system. 
The molecular coupling will be controlled by the height of the 
tunneling barrier.
To illustrate the physics we will discuss in detail quantum dot molecule
$(4,4)$ with four electrons each and contrast its properties with a 
single $N=4$ quantum dot discussed in previous section.

%
%

\subsection{One electron  spectrum of a quantum dot molecule}
\label{One_electron_spectrum}
We parameterize quantum dot molecule potential
in terms of a sum of three Gaussians   
$V(x,y)=V_L~ \exp[{-\frac{(x+a)^2+y^2}{\Delta^2}}]
        +V_R~ \exp[{-\frac{(x-a)^2+y^2}{\Delta^2}}]
+V_p \exp[{-\frac{x^2}{\Delta_{Px}^2}-\frac{y^2}{\Delta_{Py}^2}}]$.
Here $V_L,V_R$ describe the depth of the left and right quantum dot minima
located at $x=-a,y=0$ and $x=+a,y=0$, and $V_p$ is the 
central plunger gate potential controlling the tunneling barrier.
The confining potential is parameterized as $V_L=V_R=-10, \Delta=2.5, a=2$, 
and $\Delta_{Px}=0.3$, $\Delta_{Py}=2.5$, in effective atomic units.
The single particle eigenvalues and eigenfunctions are calculated
numerically by discretization of the Schr\"odinger equation
with the quantum dot molecule potential. 
The parameters of the confining potential considered in this work
are chosen to represent weakly coupled quantum dots. 
At zero magnetic field the SP levels of the electrons 
exhibit well separated S,P, and D electronic shells, and
at high magnetic field they form molecular
shells of closely spaced pairs of bonding-antibonding orbitals.
The  half-filled molecular shells correspond to electron numbers 
$(N_L=2k-1,N_R=2k-1)$ and filled shells correspond to 
$(N_L=2k,N_R=2k)$ configurations ($k$ is integer). 
The resulting single particle spectrum as a function of magnetic field 
has been presented recently in Fig. 1 of Ref.~\onlinecite{QHFerriPRL}.
In order to understand and visualize the spectrum in high magnetic field
we expand confining  potential  in the vicinity of each minimum. This gives a
parabolic potential $V(r)=m^* \omega_0^2 r^2/2$ with the strength 
$\omega_0=2\sqrt{|V_0|/\Delta^2}$.  
The low energy spectrum of each dot corresponds to two
harmonic oscillators with eigen-energies
$\epsilon_{nm}=\omega_+(n+1/2)+\omega_-(m+1/2)$.
Here  $\omega_\pm = \sqrt{\omega_0^2 + \omega_c^2/4} \pm \omega_c/2$, 
$\omega_c$ is the cyclotron energy, and
$n,m=0,1,2,...$.
With increasing magnetic field the $\omega_-$ decreases to zero
while $\omega_+$ approaches the cyclotron energy $\omega_c$, and
the states $|m,n\rangle$ evolve into the nth Landau level.
In high magnetic field the corresponding wavefunctions admit a 
description in terms of localized LLL orbitals \cite{RaminWojtekPawel}. 
In this limit linear combinations of the LLL orbitals $m$ 
from left and right dot forms molecular
shells of closely spaced symmetric-antisymmetric pairs with
eigen-energies expressed approximately as

\begin{eqnarray}
\epsilon_{m\lambda\sigma} = 
\omega_{-}(m+\frac{1}{2}) 
- \lambda \frac{\Delta_m}{2} - \frac{1}{2}\sigma \gamma \omega_{c}.
\label{energies}
\end{eqnarray}
Here $E_Z=\gamma \omega_{c}$ is the Zeeman energy, $\sigma= +1~(-1)$ 
corresponds to spin $\uparrow(\downarrow)$, $\gamma=m^* g$ and
$\lambda$ is the pseudospin index:
the symmetric (antisymmetric) orbitals are labeled by $\lambda=+1~(-1)$, 
the parity of the molecular orbitals.
$\Delta_m$ is the symmetric-antisymmetric gap.
The Zeeman coupling induces spin splitting with increasing magnetic field.
This has been illustrated in Fig. 1 of Ref.~\onlinecite{QHFerriPRL}. 
We now populate the quantum dot molecule
with $N=8$ electrons. This is an  example of 
electronic configurations corresponding to filled molecular shells.
Fig. 1 of Ref.~\onlinecite{QHFerriPRL} shows the evolution of the lowest 
energy states of noninteracting electrons, with the 
corresponding total spin $S$ states separated by vertical lines.
We find $S=0,2$, and 4 phases with even $S$ 
which correspond to simultaneous 
spin flips in each isolated dot. 

However, we also find   odd $S$ phases. 
The first odd spin state with
$S=1$ occurs between magnetic fields corresponding to 
$\omega^*_{c1} \approx 3.25$ and $\omega^*_{c2} \approx 3.9$ where
$\epsilon_{m=2,\sigma=\uparrow,\lambda=+1}=
 \epsilon_{m=1,\sigma=\downarrow,\lambda=-1}$, and
$\epsilon_{m=2,\sigma=\uparrow,\lambda=-1}=
 \epsilon_{m=1,\sigma=\downarrow,\lambda=+1}$.
The odd spin flip is related to the
splitting of energy levels due to tunneling.
Using single particle eigen-energies given by
Eq.(\ref{energies}), we find the first spin flip at the value of the magnetic 
field $B_1$

\begin{eqnarray}
\gamma \omega_c (1)  = \omega_{-}(1)- {{\Delta_2(1)+\Delta_1(1)}\over{2}} 
\label{B1}
\end{eqnarray}

\noindent
where Zeeman energy equals the single dot level splitting minus the average
symmetric-antisymmetric gap for the two levels involved. The second
spin flip takes place at $ \omega_c (2) $ such that

\begin{eqnarray}
\gamma \omega_c (2)  = \omega_{-} (2) + {{\Delta_2(2)+\Delta_1(2)}\over{2}}.
\label{B2}
\end{eqnarray}

\noindent
Hence the difference in the magnetic fields corresponding to first and second 
spin flips is a direct measure of the tunneling splitting:

\begin{eqnarray}
\gamma ( \omega_c (2) -  \omega_c (1) )  &=& { {\Delta_2(2)+\Delta_1(2)} 
\over{2} }  \nonumber \\ &&
+  { {  \Delta_2(1)+\Delta_1(1)}   \over{2} } .
\label{B3}
\end{eqnarray}

\noindent
From the spectrum of Fig. 1 of Ref.~\onlinecite{QHFerriPRL},
we observe that the states with odd spins $S=1$, and $S=3$,
are stable within narrow range of magnetic fields  
due to spin flip transitions among the electrons
that occupy the levels with energy separation proportional to
the inter-dot tunneling amplitude.
This is in contrast with the first spin flip transition in single dots
(compare with Fig. \ref{spectrum_SD}) which is stable in a wide
range of magnetic fields.
For this reason the existence of odd spin states
 in the spin phase diagram of 
quantum dot molecules can be interpreted as the measure of 
inter-dot interaction.

\subsection{Many electron quantum dot molecule spectrum}

In this section we present an analysis of spin transitions 
driven by electron-electron  interaction. We focus on the tunnel coupled 
lowest Landau level
orbitals $m$.
Denoting the creation (annihilation) operators for electron in
non-interacting SP state $|m\lambda\sigma\rangle$ by
$c^\dagger_{m\lambda\sigma}~(c_{m\lambda\sigma})$ (with $\sigma$
as spin label), the Hamiltonian of an interacting system, Eq. \ref{ham1},
can be written as
\begin{eqnarray}
H&=&\sum_{m\lambda}\sum_{\sigma}
\epsilon_{m\lambda\sigma}
c^\dagger_{m\lambda\sigma} c_{m\lambda\sigma}
\nonumber \\ &&
+ \frac{1}{2}\sum_{\{m,\lambda\}}\sum_{\sigma\sigma'}
\langle m_1\lambda_1\sigma,m_2\lambda_2\sigma' |
V |m_3\lambda_3\sigma',m_4\lambda_4\sigma \rangle \nonumber \\ &&
\times
c^\dagger_{m_1\lambda_1\sigma} 
c^\dagger_{m_2\lambda_2\sigma'}
c_{m_3\lambda_3\sigma'} 
c_{m_4\lambda_4\sigma}
\label{multiparticle}
\end{eqnarray}
The single particle states of coupled quantum dot molecules 
in magnetic field are labeled by the 
orbital quantum numbers $m$, 
and the pseudospin index $\lambda$.
The first term in Eq. \ref{multiparticle} is the single particle 
Hamiltonian, and $V_{\alpha\beta\mu\nu}$
is the two-body Coulomb matrix element.
Here $\{\alpha, \beta, \mu, \nu\}$ represent the states with 
quantum numbers $(m,\lambda,\sigma)$.
We now turn to the construction of the relevant configurations.
\subsection{$S=0$, $\nu=2$ state}
\begin{figure}
\begin{center}
\includegraphics[width=0.8\linewidth]{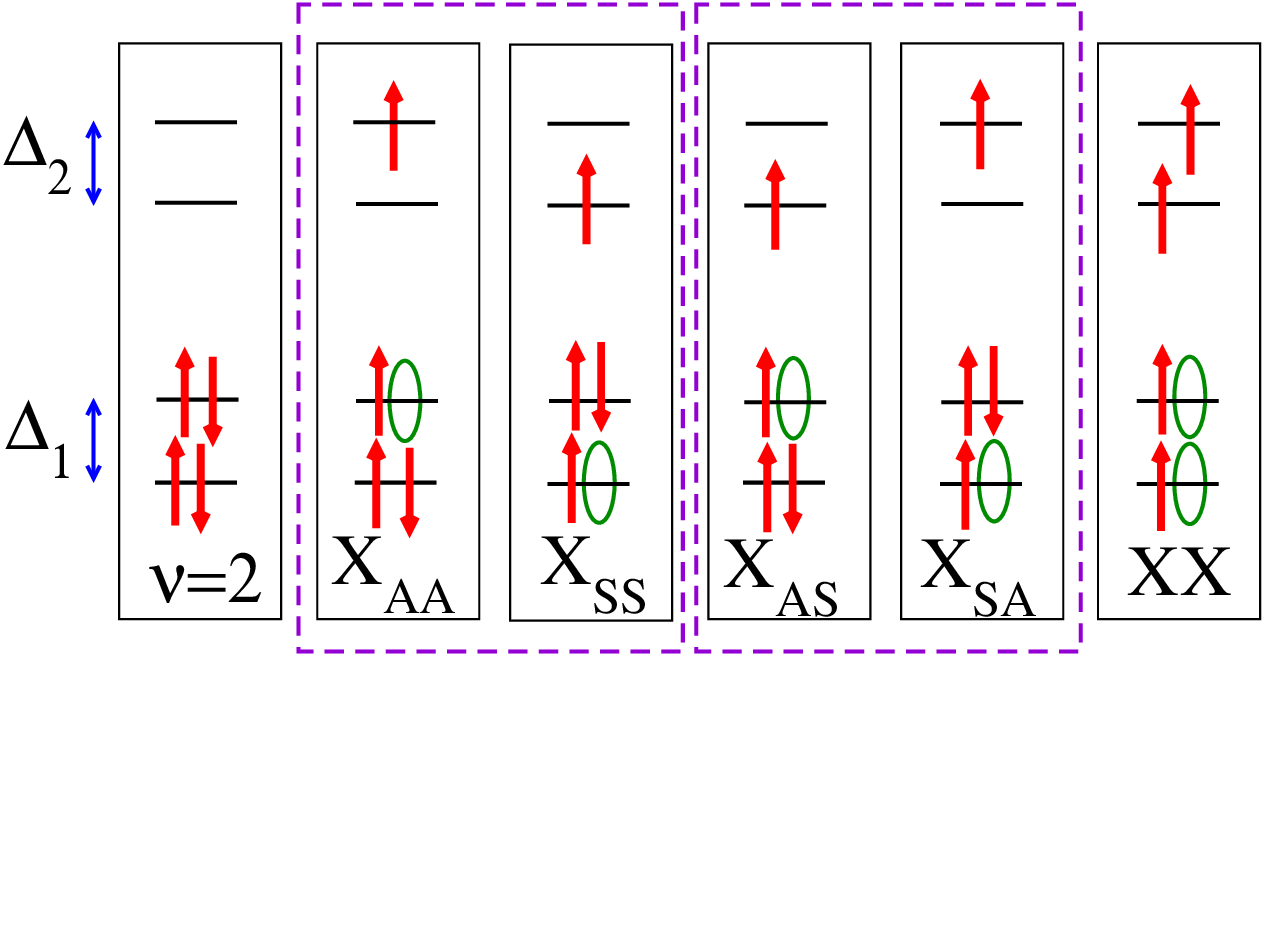}\vspace{-1cm}
\noindent
\caption{(Color online) The basis of spin configurations in high magnetic fields.
The first spin transition states $S=1$ identify with two independent
set: $\{X_{SS},X_{AA}\}$, and $\{X_{SA},X_{AS}\}$.
In the former the electron-hole transitions occurs between the
states with the same symmetry and hence
they do not mix with the latter which exhibit the process of 
electron-hole excitations between states with opposite parity. 
}
\label{Xsa}
\end{center}
\end{figure}

The $\nu=2$  state of quantum dot molecule with $N$
electrons and total spin $S=0$, shown in Fig. \ref{Xsa}, is the 
product of electron creation operators
\begin{eqnarray}
|\nu=2\rangle = \prod_{m=0}^{N/4-1}\prod_{\lambda =1,2}
\prod_{\sigma=\uparrow,\downarrow} c^\dagger_{m \lambda \sigma}|0\rangle
\end{eqnarray}
The energy associated with this state 
\begin{eqnarray}
E_{\nu=2}= \sum_{m=0}^{N/4 -1} \sum_{\lambda =\pm 1} \sum_{\sigma}
\left[\epsilon_{m\lambda\sigma} + \Sigma(m,\lambda,\sigma)\right]
\end{eqnarray}
can be expressed in terms of self-energy $\Sigma(m,\lambda,\sigma)$ 
\begin{eqnarray}
\Sigma(m,\lambda,\sigma) &=& 
\sum_{m'=0}^{N/4 -1} \sum_{\lambda'=\pm 1} 
(2\langle m \lambda, m' \lambda'|V|m' \lambda', m \lambda\rangle 
\nonumber \\ &&
- \langle m \lambda, m' \lambda'|V|m \lambda, m' \lambda'\rangle).
\end{eqnarray}

\subsection{$S=1$ spin exciton}
The $S=1$ spin flip excitation is constructed by removing an electron
from occupied $j$ orbital in a  $\nu=2$ state, and putting it into an unoccupied state $i$,
\begin{eqnarray}
|X_{j\rightarrow i}\rangle = c^\dagger_i c_j |\nu=2 \rangle ,
\end{eqnarray}
where $j\equiv (m,\lambda,\downarrow)$, and $i\equiv (m',\lambda',\uparrow)$.
Denoting quasi-particle energy levels (electrons dressed by interaction)
by $\varepsilon_i = \epsilon_i + \Sigma(i)$
the energy of one exciton follows
\begin{eqnarray}
\Delta E_{X_{j\rightarrow i}} = \varepsilon_i - \varepsilon_j 
- \langle i,j|V|j,i\rangle 
\label{EX}
\end{eqnarray}
where $\Delta E_{X_{j\rightarrow i}} = E_{X_{j\rightarrow i}}- E_{\nu=2}$
is the energy of exciton relative to the $\nu=2$ state energy.
The last term in Eq. (\ref{EX}) is the electron-hole Coulomb  
attraction.
The lowest energy states of the single exciton
corresponding to the first spin flip state are depicted in Fig. \ref{Xsa}. 
We classify the single excitons based on their relative parity with
respect to the parity of the $\nu=2$ state.
Pair of states  $(X_{SS}, X_{AA})$ describes transitions between pairs of
levels with the same parity and so parity is conserved by these 
transitions.
In contrast, spin flip transitions represented by $(X_{SA}, X_{AS})$
do not conserve parity as they describe transitions between
pairs of levels with opposite parity. 
We identify $(X_{SS}, X_{AA})$, and $(X_{SA}, X_{AS})$ by 
their parity quantum numbers $\pi=+1$ and $\pi=-1$ ,
respectively.
It is important to note that Coulomb interactions do not mix 
excitons with different parities, and the CI Hamiltonian 
constructed in the basis of spin excitons is block diagonal.
These pairs of independent excitons are 
shown inside the boxes in Fig. \ref{Xsa}.

\begin{figure}
\begin{center}
\includegraphics[width=0.8\linewidth]{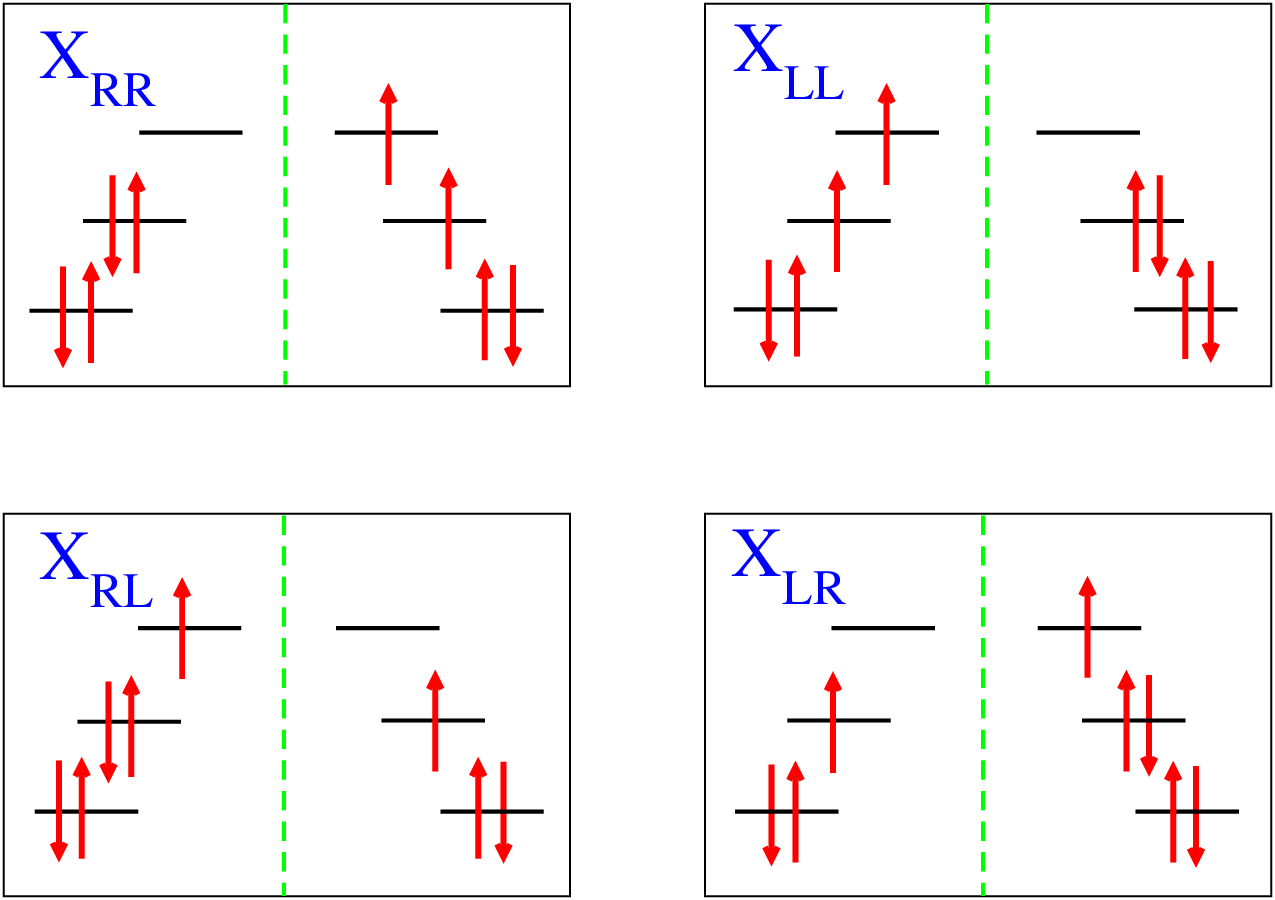}
\noindent
\caption{
(Color online) The lowest energy excitons in real space
$|X_{RR}\rangle=|X_{(1,R,\downarrow)
	\rightarrow (2,R,\uparrow)}\rangle$,
$|X_{LL}\rangle=|X_{(1,L,\downarrow)
	\rightarrow (2,L,\uparrow)}\rangle$,
$|X_{RL}\rangle=|X_{(1,R,\downarrow)
	\rightarrow (2,L,\uparrow)}\rangle$,
and
$|X_{LR}\rangle=|X_{(1,L,\downarrow)
	\rightarrow (2,R,\uparrow)}\rangle$.
The magnetic ordering of these states can be described by
means of ferrimagnetic coupling $(S_L=0,S_R=1)$ for $X_{RR}$,
and $(S_L=1,S_R=0)$ for $X_{LL}$,
and ferromagnetic coupling $(S_L=1/2,S_R=1/2)$ for $X_{RL}$,
and $X_{LR}$.
}
\label{X}
\end{center}
\end{figure}

In real space, single excitons can be expressed in terms of linear
combination of excitons localized in each dot.
This basis is shown in Fig. \ref{X}.
In each isolated dot, at a critical field, a transition from the $S=0$ 
singlet to the $S=1$ triplet state takes place.
This configuration, corresponding to $X_{RR}$ and $X_{LL}$
in Fig. \ref{X}, is equivalent to a localized 
electron-hole excitation in one dot in the presence of 
background of electrons in the other dot.
Because of the geometrical symmetry associated with the electron-hole 
excitation, the ground state of the system without  external
bias and interdot tunneling has double degeneracy:
The state with $(S_L=0,S_R=1)$ has exactly the same energy as the
state $(S_L=1,S_R=0)$.
The many body wavefunction of such a molecular state
can be expressed as linear combination of 
degenerate states $|S_L=0,S_R=1\rangle$ and $|S_L=1,S_R=0\rangle$.
We identify these pairs of molecular states with  quantum Hall ferrimagnets.
For a range of magnetic field these two states are separated
from another pair of single excitations $X_{LR}$ and $X_{RL}$ 
with $(S_L=1/2,S_R=1/2)$ by an energy gap due to
Coulomb interactions.
In general, in the case of filled shells, molecular states with 
odd total spin $S$ correspond to quantum Hall ferrimagnets.
Our analysis of spin transitions in real space is presented in 
section \ref{realspace}.

\subsection{$S=2$ spin bi-exciton}
With increasing magnetic fields,
a higher polarized state with $S=2$,
$(S_L=1,S_R=1)$ equivalent to a spin bi-exciton state
appears.
A bi-exciton is constructed by removing a pair of electrons
from occupied states and putting them into unoccupied states
(see Fig. \ref{Xsa}),
\begin{eqnarray}
|XX_{j\rightarrow i, k\rightarrow l}\rangle 
= c^\dagger_i c^\dagger_l c_j c_k | \nu=2 \rangle.
\end{eqnarray}
The energy of biexciton can be decomposed into the
energy of two single excitons plus their interaction
\begin{eqnarray}
\Delta E_{XX_{j\rightarrow i, k\rightarrow l}} = 
\Delta E_{X_{j\rightarrow i}} + \Delta E_{X_{k\rightarrow l}}
+ \delta V,
\label{EXX}
\end{eqnarray}
where 
$\Delta E_{X_{j\rightarrow i}}$ has been defined in Eq. \ref{EX}.
$\delta V$ is the binding energy between two excitons,
accounting for the electron-electron, electron-hole and hole-hole
interactions
\begin{eqnarray}
\delta V &=& \langle l,i |V| i,l \rangle - \langle l,i |V| l,i \rangle 
           \nonumber\\&&
           - \langle l,j |V| j,l \rangle                                 
           - \langle k,i |V| i,k \rangle                                 
	   \nonumber\\&&
           + \langle j,k |V| k,j \rangle - \langle j,k |V| j,k \rangle.   
\end{eqnarray}

\subsection{First versus second spin flip in a quantum dot molecule}
\label{ExcitonicCondensation}

\begin{figure}
\begin{center}
\includegraphics[width=0.8\linewidth]{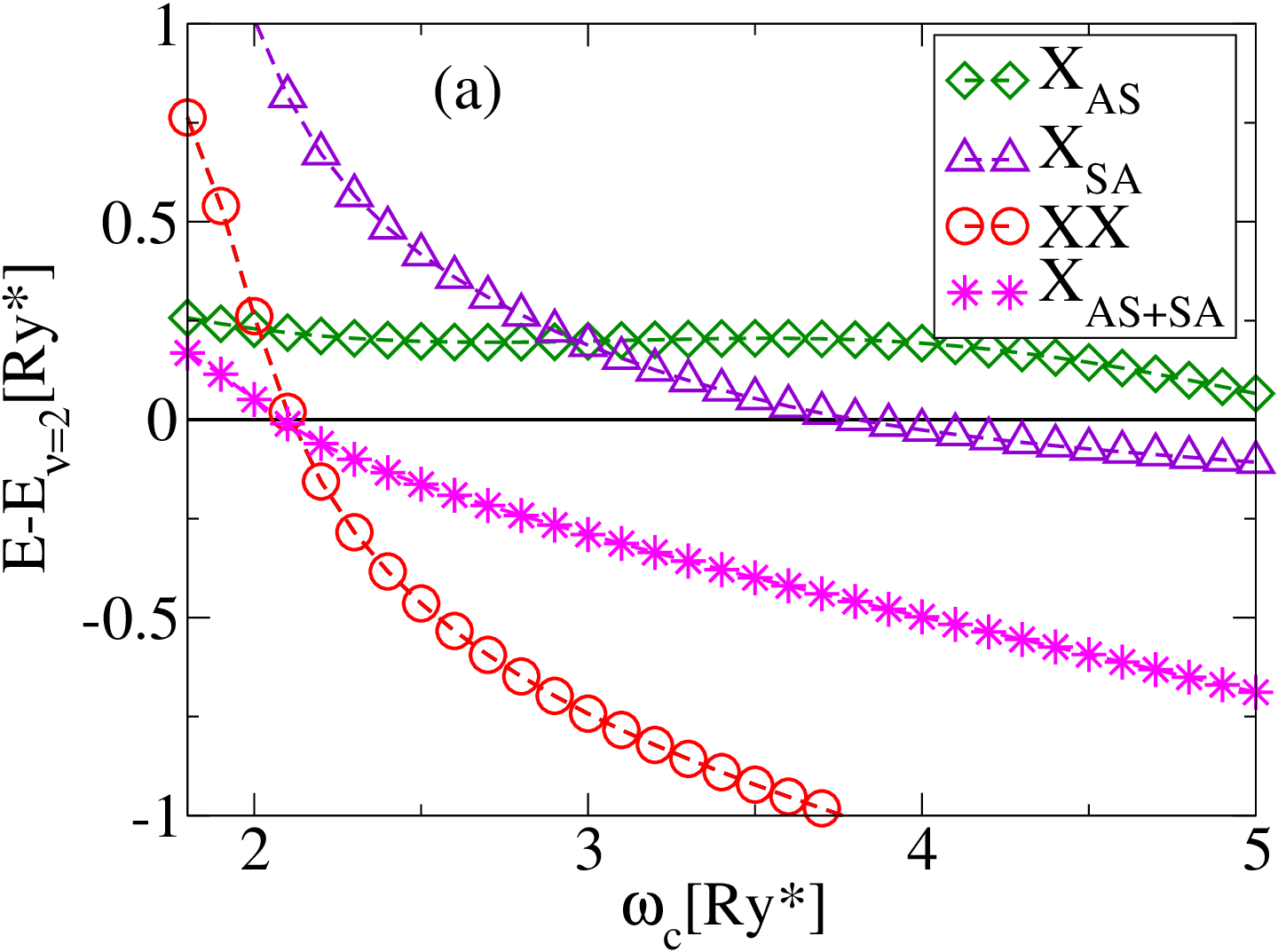}\vspace{1cm}
\includegraphics[width=0.8\linewidth]{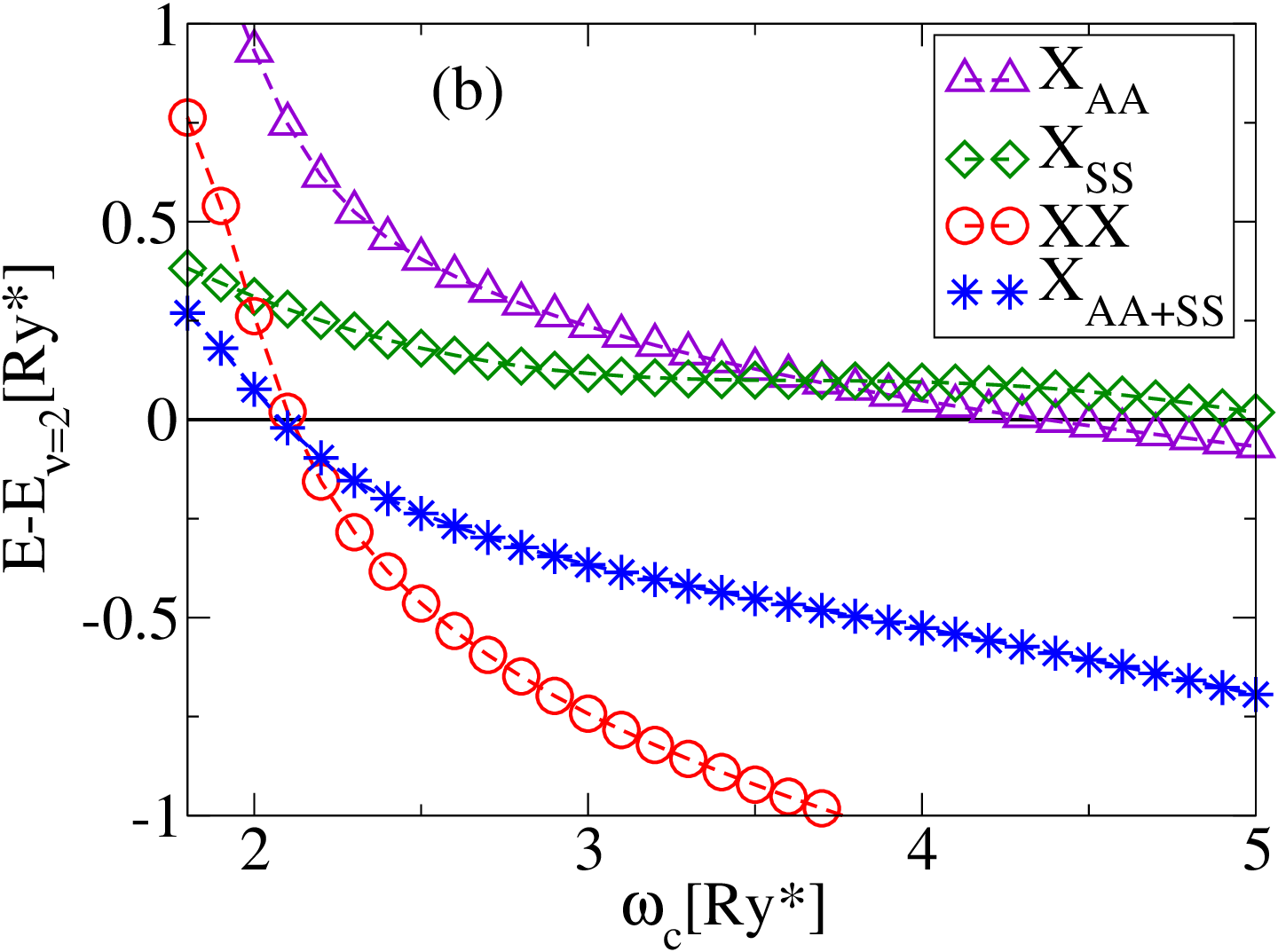}\vspace{1cm}
\noindent
\caption{(Color online) 
(a) The energies of two  single spin excitons $X_{SA}$ and $X_{AS}$ 
with odd parity, the energy of the odd parity correlated exciton 
$X_{SA+AS}$, and the energy of the spin bi-exciton $S=2$ state as a 
function of magnetic field. All energies measured from the energy of the 
$\nu=2,S=0$ state.
(b) The energies of two  single spin excitons $X_{SS}$ and $X_{AA}$ with even 
parity, the energy of the even parity correlated exciton $X_{SS+AA}$, 
and the energy of the spin bi-exciton $S=2$ state as a function of magnetic 
field. 
All energies measured from the energy of the $\nu=2,S=0$ state.
}
\label{T1_LR}
\end{center}
\end{figure}

Unlike in a single quantum dot, first spin flip state  $S=1$ 
is not an eigenstate of the coupled quantum dot Hamiltonian. 
There are two possible spin excitons for a given parity,
and they are coupled by Coulomb interactions. 
We use two distinct
single exciton basis $\{X_{SS},X_{AA}\}$ and $\{X_{SA},X_{AS}\}$, 
labeled by parity $\pi=+1$ and $\pi=-1$, to
construct the two  $2\times 2$ Hamiltonians.
We denote by $\Delta E_{X_{SS}+X_{AA}}$, $\Delta E_{X_{SA}+X_{AS}}$ 
the lowest eigen-energies of these Hamiltonians.

Fig. \ref{T1_LR}(a) shows the numerically calculated 
energies of odd parity spin excitons as a function of magnetic field.
The energy $\Delta E_{X_{AS}}$  of spin exciton  $X_{AS}$ is positive 
for magnetic fields shown
but the energy $\Delta E_{X_{SA}}$ of spin exciton $X_{SA}$ becomes negative 
at $\omega_c=3.8$ i.e.
the  $X_{SA}$ spin flip state becomes the lower energy state than the 
$\nu=2, S=0$  state. 
However, in stark contrast with a single quantum dot,
Fig. \ref{SD_SP_CI}, we find that the second spin flip state $XX$ becomes 
the ground state
at lower magnetic field $\omega_c=2.1$. Hence, unlike in a single quantum dot
we find a transition from spin singlet $S=0$ state directly to $S=2$ second 
spin flip state.
This is a transition corresponding to even total spin numbers, as if the two 
dots were flipping their
spin simultaneously. However, correlations or mixing of the two spin excitons 
$X_{AS}$ and
$X_{SA}$ lowers the energy of the spin exciton. The energy 
$\Delta E_{X_{SA}+X_{AS}}$) of the correlated 
single spin state is significantly lower and equals both the energy 
$\Delta E_{XX}$ of the second spin flip bi-exciton and of  the $\nu=2, S=0$  
state at $\omega_c=2.1$. At this value of the magnetic field
the energy of the bi-exciton and of the exciton are almost identical, 
and we might expect that for
larger number of configurations correlations will stabilize the single 
spin flip exciton even further.
The effect of correlations on the even parity excitons $X_{AA}$ and $X_{SS}$ 
is shown in 
Fig. \ref{T1_LR}(b). We see that mixing of the two even parity excitons lowers 
the energy $\Delta E_{X_{SS}+X_{AA}}$ of the correlated 
single spin state. This energy   equals both the energy $\Delta E_{XX}$ of 
the second spin 
flip bi-exciton and of  the $\nu=2, S=0$  state at $\omega_c=2.1$. 
By comparison with Fig. \ref{T1_LR}(a) we see that 
the value of $\omega_c=2.1$ also corresponds to the change in parity of 
the single spin exciton $S=1$ state.

As illustrated in Fig. \ref{T1_LR}, 
with  mixing of single exciton configurations
the energy of quantum dot molecule exhibits four-fold degeneracy
at $\omega^*_{c1}$ where $S=0$,  $S=2$ and two different parity $S=1$ 
states become
the lowest energy states.
The $S=1$ states show stability in a narrow range of magnetic
field, within the accuracy of our numerical results.
With further increase of magnetic field, 
single excitons condense into pairs of excitons forming bi-excitons.
The existence of single, odd, spin excitons is hence a signature of 
electronic correlations. 
These states do not exist at the Hartree-Fock level.

\begin{figure}
\begin{center}
\includegraphics[width=0.8\linewidth]{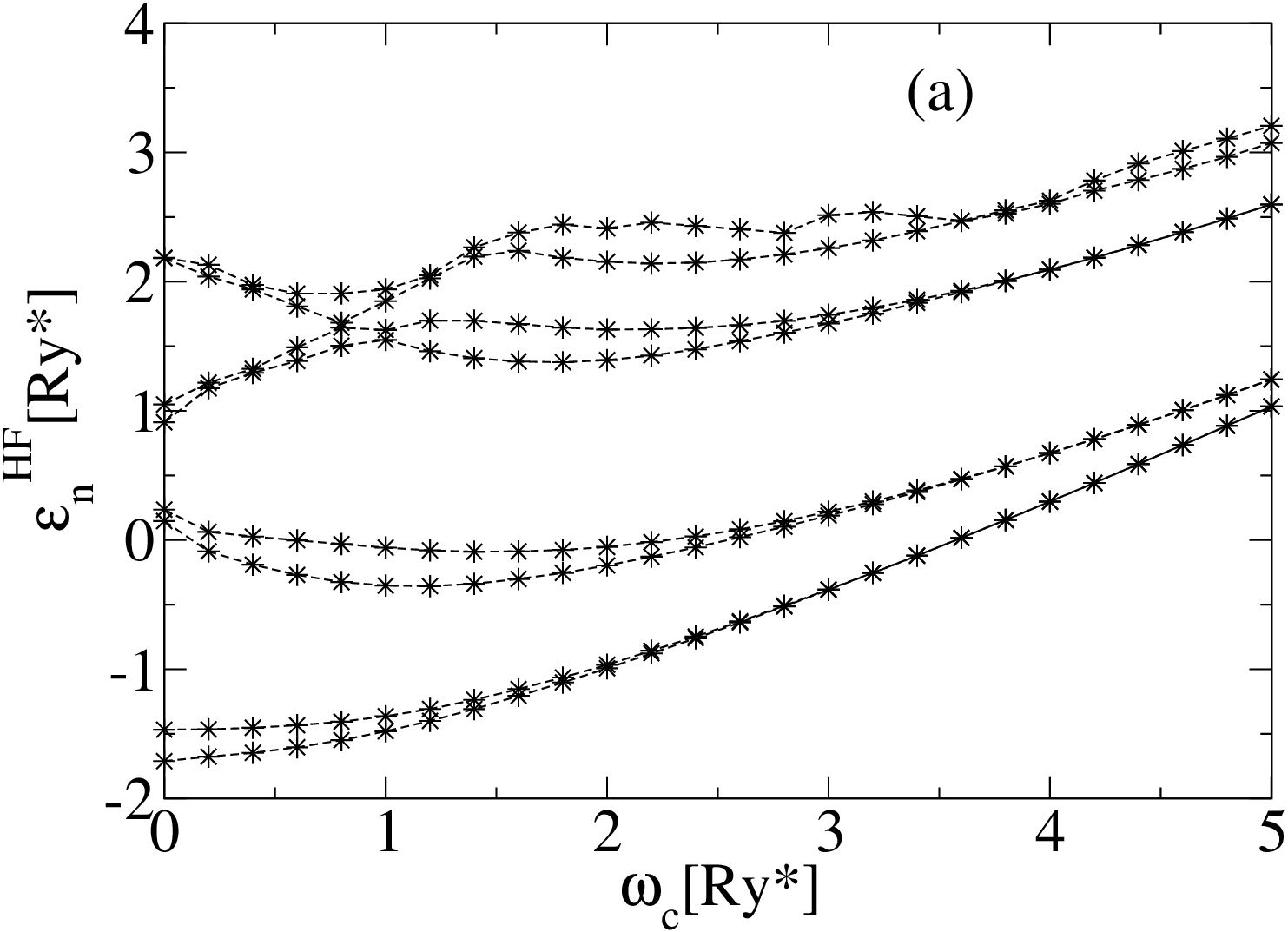}\vspace{1cm}
\includegraphics[width=0.8\linewidth]{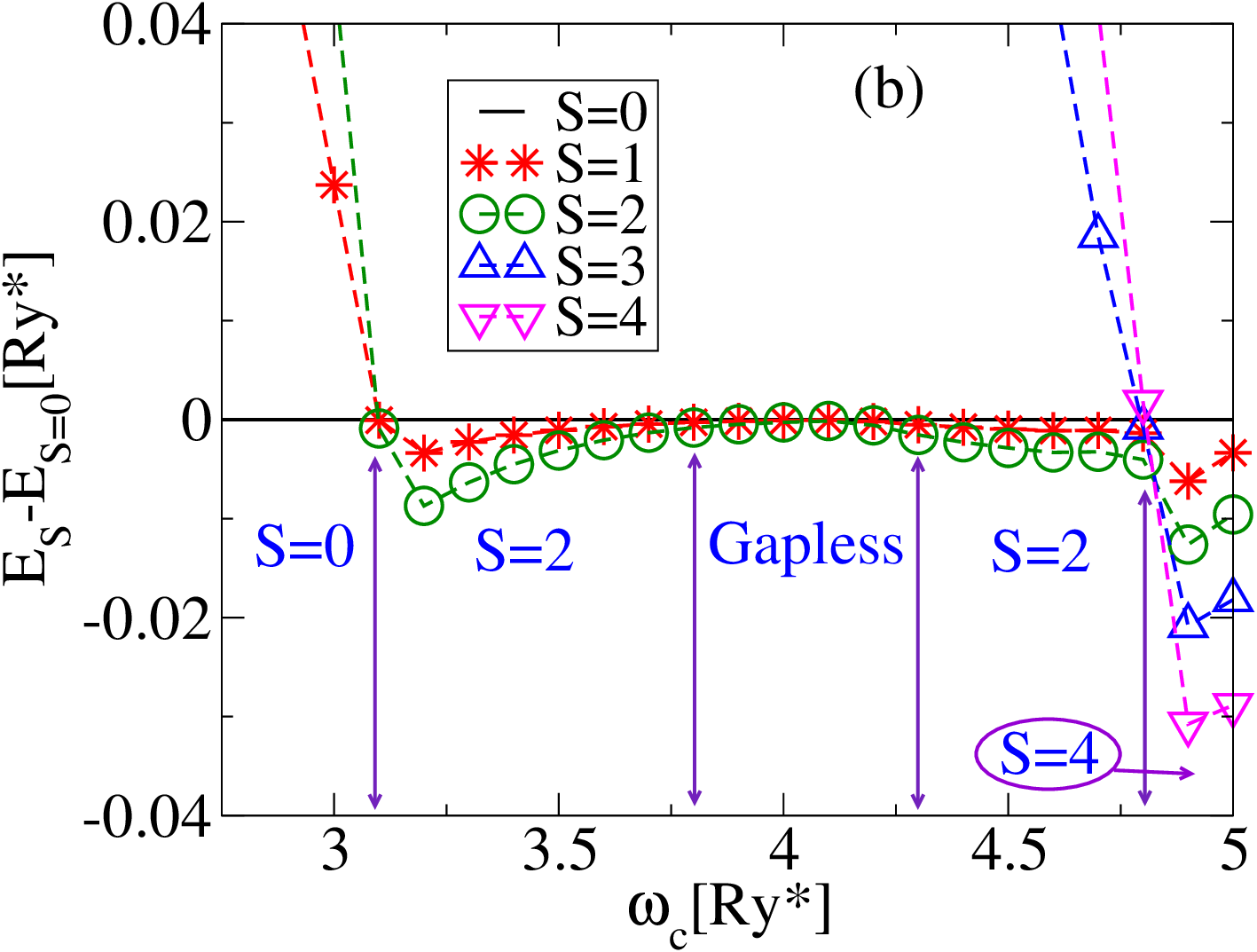}
\noindent
\caption{(Color online) 
(a) URHF energies vs. cyclotron energy for $N=8$ electrons
with $S_z=0$.
(b) Evolution of lowest energies for $S=0,1,2$ states  calculated by
URHF-CI method. The energies are measured from the energy of the  $S=0$ state.
} 
\label{HF8e}
\end{center}
\end{figure}

To support the assertion that correlations are responsible
for the existence of odd spin excitons we employ URHF-CI calculation.
From the solution of the Pople-Nesbet equations 
we obtain HF eigen-energies,  shown in Fig. \ref{HF8e} (a), 
for $N=8$ electrons with $S_z=0$, as a function of the magnetic field. 
The HF wavefunctions are used as a basis
in HF-CI calculation of the ground state.  We employ
$N_s=8$ HF basis states (equivalent to $N_C=4900$ configurations) to
calculate the total spin of electrons in quantum dot molecule as a function
of the magnetic field.
From this calculation we find that the $S=0$, $S=1$, and
$S=2$ states are almost degenerate at $\omega^*_{c1}=3.1$.
The prediction of URHF-CI is
in qualitative agreement with effective SP-CI model, presented above.
The direct, exchange, and correlation energies calculated
by URHF-CI shift the transition point to  higher magnetic fields.

\begin{figure}
\begin{center}
\includegraphics[width=0.8\linewidth]{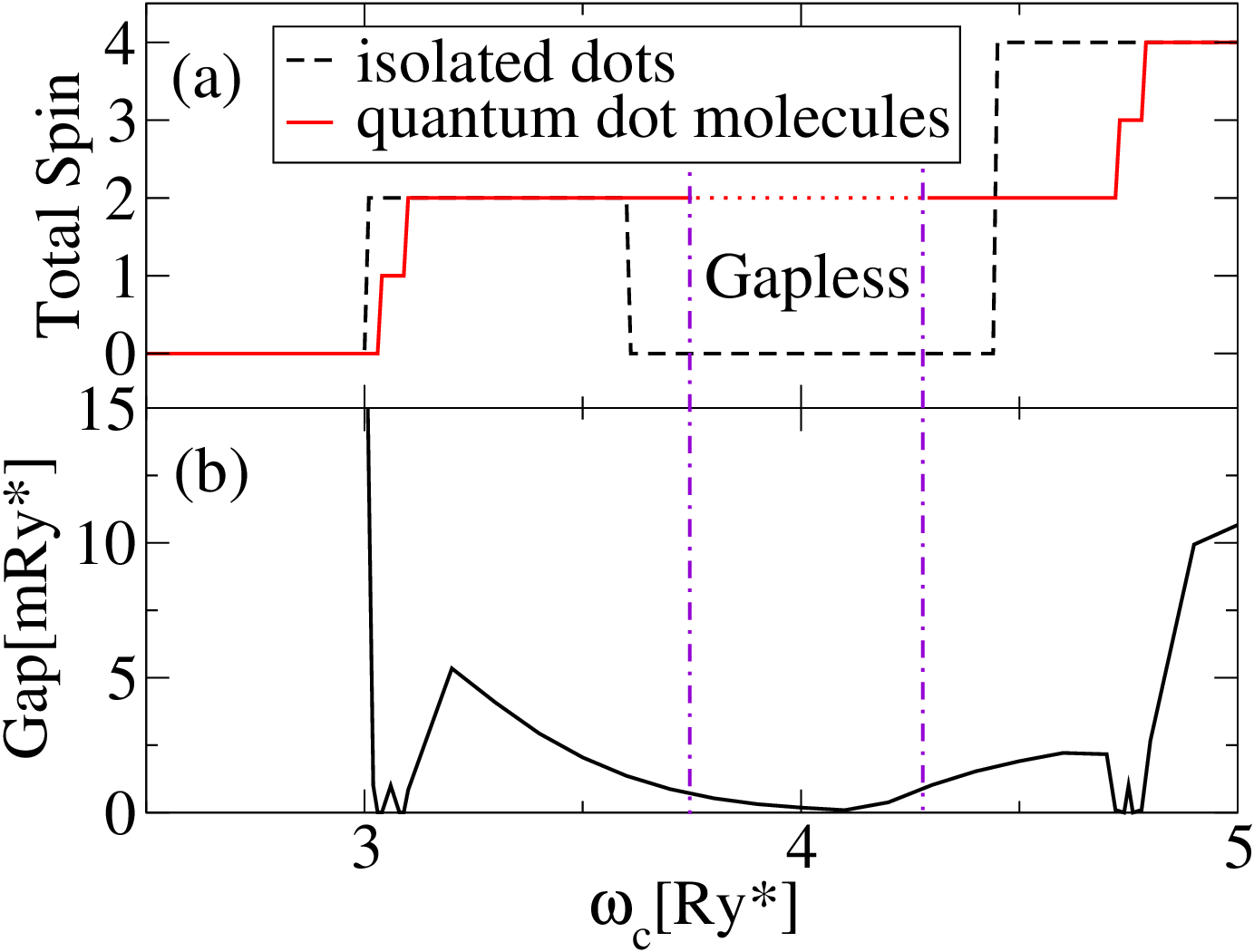}\vspace{1cm}
\noindent
\caption{(Color online) 
(a) The evolution of spin  of the $N=8$ electron 
quantum dot molecules as a function of magnetic field for $g=0$ and $V_p=7$ 
(solid line). 
For comparison the spin evolution of two isolated dots is shown (dashed line).
The width of odd spin plateaux
have been artificially enlarged to be visible by eyes.
The vertical dashed line (with purple-color online) tends to show 
qualitatively a range of $\omega_c$ in which the gap vanishes. 
The horizontal dashed line (with red-color online) shows the corresponding 
spin one state with zero gap.
(b) The evolution of the energy gap  of the 
quantum dot molecules as a function of magnetic field.
} 
\label{Gap1}
\end{center}
\end{figure}

With  increasing of the magnetic field   we find that the gap between different total
spin states tends to vanish. This gapless phase is seen in 
Figs. (\ref{HF8e}) and (\ref{Gap1})
in the vicinity of  $\omega_c=4$. This phase is
followed by the $S=2$, $S=3$ and $S=4$ states.
The latter state corresponds to a fully spin polarized $\nu=1$ droplet.

Let us now summarize  the similarities and the differences 
in the evolution of total  spin  of two isolated dots and
a quantum dot molecule.
Fig. \ref{Gap1} (a) shows the evolution of total spin with 
increasing magnetic field
for two noninteracting $N=4$ quantum dots and for $N=8$ quantum dot molecule. 
Fig.\ref{Gap1}(b)
shows the energy gap of the molecule as a function of magnetic field.
The noninteracting quantum dots spin evolution is obtained by adding results 
from two isolated dots,
each dot evolving with magnetic field according to Fig. \ref{SD_HFCI}.
We find that the effect of interdot interaction is to renormalize the 
magnetic fields at which spin transitions take place, and more importantly,
to lead to the appearance of odd spin states $S=1$ and $S=3$. While the 
existence of odd spin states
is most striking, the presence of spin polarized phases is also nontrivial. 
The fact that
spins of electrons on two quantum dots align demonstrates the existence of 
ferromagnetic 
dot-dot coupling. In the case of antiferromagnetic coupling there would have 
been no
net spin eventhough each dot has finite spin. We will show in the next
section that such antiferromagnetic coupling exists in the $N=8$ molecule at 
low magnetic fields.

\begin{figure}
\begin{center}
\includegraphics[width=0.8\linewidth]{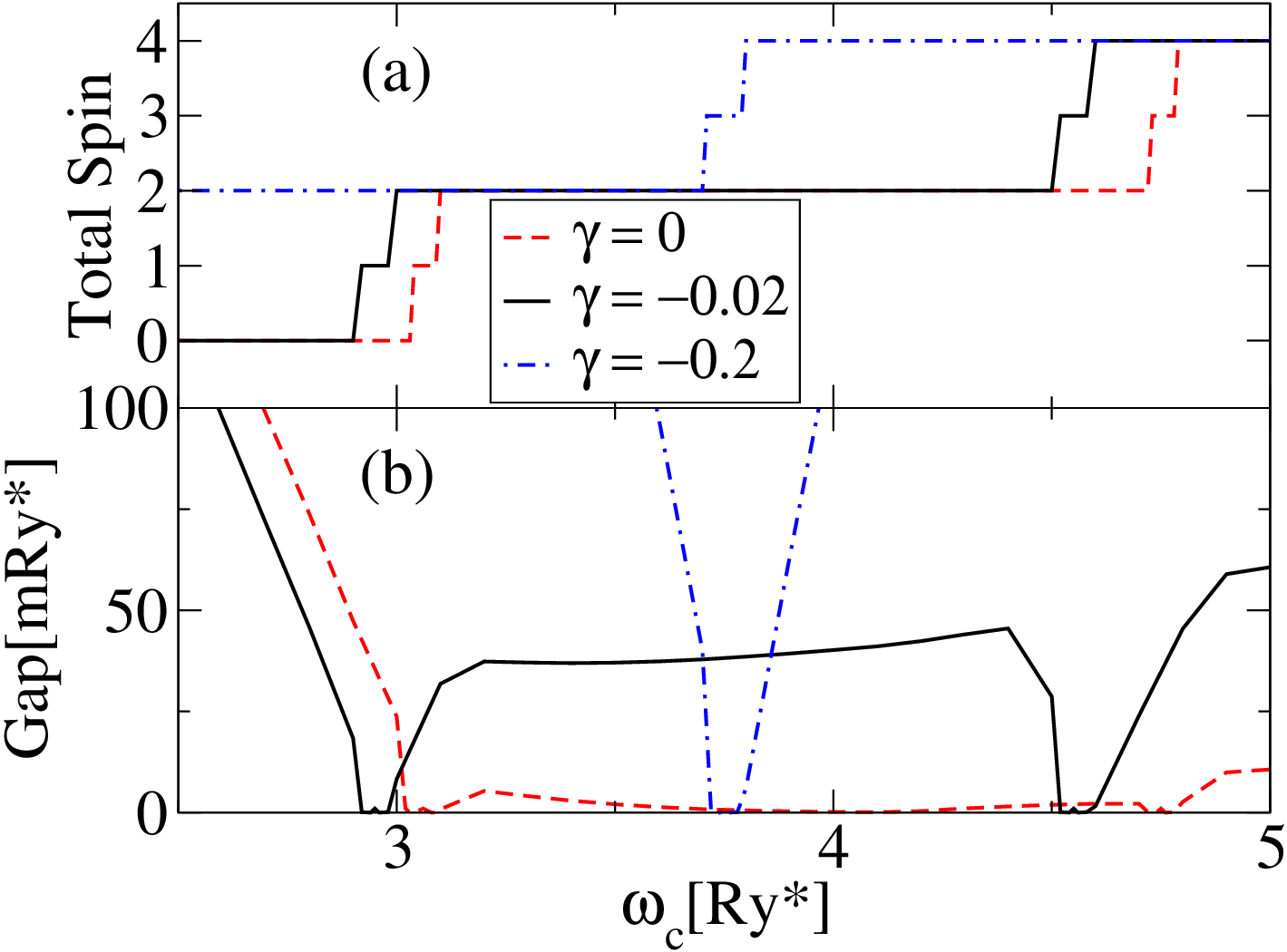}\vspace{1cm}
\noindent
\caption{
(Color online) The effect of increasing Zeeman energy on the  (a)  evolution of spin  
and (b) evolution of energy gap
of the $N=8$ electron quantum dot molecules as a function of magnetic field. 
The width of $g=0$ odd spin plateaux
have been artificially enlarged to be visible by eyes.
} 
\label{Gap2}
\end{center}
\end{figure}

Finally, following the gap in Fig.\ref{Gap1}(b) we find the existence of low 
spin ``gapless" states,
molecular analogs of spin singlet bi-excitons.
Hence Fig.\ref{Gap1} shows that electronic inter-dot correlations 
stabilize the odd spin 
phases but their stability range is very narrow. As shown in 
Sec. (\ref{One_electron_spectrum}) without electron-electron interactions 
competition between 
quantum mechanical tunneling and Zeeman energy was responsible for the 
existence
of odd spin phases. The effect of finite Zeeman energy is similar in an 
interacting system.
Fig.\ref{Gap2} shows the effect of increasing Zeeman energy on the 
evolution of spin  
and  energy gap
of the $N=8$ electron quantum dot molecules as a function of magnetic field.
All parameters are the same as in Fig.\ref{Gap1}. We see that increasing Zeeman energy
renormalizes  the magnetic field value of spin flips and, more importantly,
 stabilizes the odd spin phases.
From the evolution of the energy gap shown in Fig.\ref{Gap2}b  we also see the 
vanishing of
the low spin depolarized phase in the vicinity of $\omega_c=4$ and the 
stabilization of the
spin polarized $S=2$ phase. 
The Zeeman energy depends on the g-factor. 
For GaAs the g-factor 
is $g_{GaAs}=-0.44$ while for InAs and InSb the g-factors are 
$g_{InAs}=-14$, and $g_{InSb}=-50$. 
\cite{GuyPhysicaB,Kouwenhoven2,Hanson03,GuyPRB,McCombe,IgorRMP}
Hence by adding 
In one can hope to tune the g-factor of quantum dot molecules.

\begin{figure}
\begin{center}
\includegraphics[width=0.8\linewidth]{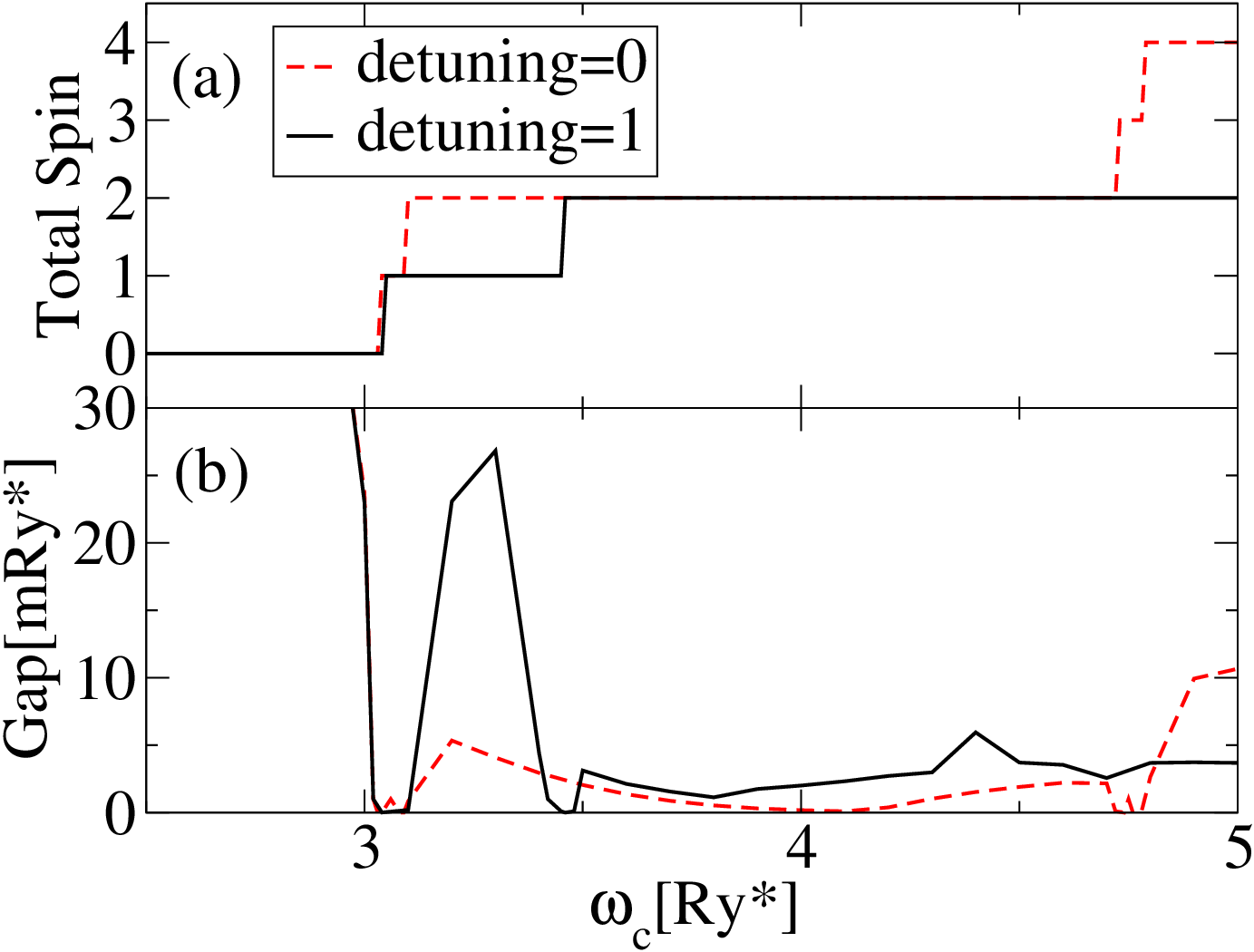}\vspace{1cm}
\noindent
\caption{(Color online) 
The effect of detuning, i.e., difference in confining potential between left dot
$V_L$ and right dot $V_R$ on the  (a)  evolution of spin  
and (b) evolution of energy gap
of the $N=8$ electron quantum dot molecules as a function of magnetic field. 
} 
\label{Gap3}
\end{center}
\end{figure}

To conclude our analysis of the $N=8$ electron quantum dot molecule we 
discuss the effect
of asymmetry between the two dots. While for molecules built out of two 
atoms each component is
identical, quantum dots are defined by gates or etching and one must 
understand the effect of
differences between the two dots on the stability diagram \cite{MichRamin}. 
In Fig.\ref{Gap3} we show the evolution of spin  
and  energy gap
of the $N=8$ electron quantum dot molecules as a function of magnetic field. 
The two dots are different,
with confining potential of the left dot $V_L=-10$ unchanged but the potential 
of the right dot detuned by
$1Ry$ to $V_R=-11$. 
As anticipated, the effect of detuning results in increased stability of the 
odd spin flip state $S=1$. 


\section{Real space analysis of spin transitions in quantum 
dot molecules}
\label{realspace}

In this section, we describe the analysis of spin transitions in real space. 
Eq.~(\ref{multiparticle}) describes 
the Hamiltonian of an interacting system in second quantization
in non-interacting SP state $|m\lambda\sigma\rangle$.
The single particle states of coupled quantum dot molecules 
in magnetic field are labeled by the single dot
orbital quantum numbers $m$, 
and the pseudospin index $\lambda$.
The symmetric (antisymmetric) orbitals are labeled by $\lambda=1~(-1)$, 
the parity of the orbitals in two symmetric dots.
$m$ represents the combined Landau level, 
and angular momentum quantum numbers, $m\equiv (n,l)$.
The first term in Eq. (\ref{multiparticle}) is the single particle Hamiltonian,
and $V_{\alpha\sigma,\beta\sigma',\mu\sigma',\nu\sigma} =
\int d\vec{r} \int d\vec{r'}
\tilde{\varphi}^*_{\alpha\sigma}(\vec{r})
\tilde{\varphi}^*_{\beta\sigma'}(\vec{r'})
\frac{e^2}{\epsilon|\vec{r}-\vec{r'}|}
\tilde{\varphi}_{\mu\sigma'}(\vec{r'})
\tilde{\varphi}_{\nu\sigma}(\vec{r})$,
is the two-body Coulomb matrix element.
Here $\{\alpha, \beta, \mu, \nu\}$ represent the states with 
orbital quantum numbers $(m,\lambda)$.

Alternatively
denoting the creation (annihilation) operators for electron in
non-interacting localized SP state $|m s\sigma\rangle$ by
$d^\dagger_{m s\sigma}~(d_{m s\sigma})$,
the Hamiltonian of an interacting system in second quantization
can be written as
\begin{eqnarray}
H&=&\sum_{m s}\sum_{\sigma}
\tilde{\epsilon}_{m s}
d^\dagger_{m s\sigma} d_{m s\sigma}
\nonumber \\ &&
+ \sum_{m \sigma} t_m \sum_{s_1 s_2} (1-\delta_{s_1 s_2})
d^\dagger_{m s_1\sigma} d_{m s_2\sigma}
\nonumber \\ &&
+ \frac{1}{2}\sum_{\{m, s\}}\sum_{\sigma\sigma'}
\langle m_1 s_1\sigma,m_2 s_2\sigma' | V |m_3 s_3\sigma',m_4 s_4\sigma \rangle \nonumber \\ &&
\times
d^\dagger_{m_1 s_1\sigma} 
d^\dagger_{m_2 s_2\sigma'}
d_{m_3 s_3\sigma'} 
d_{m_4 s_4\sigma}
\label{multiparticle_1}
\end{eqnarray}
Here $s=1~(2)$ are pseudospin labels of electron 
localized in left (right) dot.
The relation between Eqs. \ref{multiparticle}, and \ref{multiparticle_1} 
can be established by a rotation in pseudospin space
$c^\dagger_{m\lambda\sigma} = \frac{1}{\sqrt{2}}\sum_{s=1}^2
\lambda^{s-1} d^\dagger_{m s \sigma}$.
We find $\tilde{\epsilon}_{m s} = 
(\epsilon_{m,\lambda=1} + \epsilon_{m,\lambda=-1})/2$, 
$t_m = (\epsilon_{m,\lambda=1} - \epsilon_{m,\lambda=-1})/2$, and
\begin{eqnarray}
&&\langle m_1 s_1\sigma,m_2 s_2\sigma' |
V |m_3 s_3\sigma',m_4 s_4\sigma \rangle 
= \frac{1}{4}\sum_{\lambda}
\lambda_1^{s_1 - 1} \lambda_2^{s_2 - 1} \nonumber \\ &&
\lambda_3^{s_3 - 1} \lambda_4^{s_4 - 1}
\langle m_1\lambda_1\sigma,m_2\lambda_2\sigma' |
V |m_3\lambda_3\sigma',m_4\lambda_4\sigma \rangle 
\end{eqnarray}

\subsection{$S=0$ ground state}
The $\nu=2$ state of quantum dot molecule with $N$
electrons and total spin $S=0$ is the 
product of spin polarized localized electrons 
\begin{eqnarray}
|\nu=2\rangle = \prod_{m=0}^{N/4-1}\prod_{s=1,2}
\prod_{\sigma=\uparrow,\downarrow} d^\dagger_{ms\sigma}|0\rangle
\end{eqnarray}
The energy associated with this state follows 
\begin{eqnarray}
E_{\nu=2}= \sum_{m=0}^{N/4 -1} \sum_{s=1}^2 
\left(2\tilde{\epsilon}_{ms} + \Sigma(m,s)\right)
\end{eqnarray}
where $\Sigma(m,s)$ is the electron self-energy:
\begin{eqnarray}
\Sigma(m,s) &=& \sum_{m'=0}^{N/4 -1} \sum_{s'=1}^2 
(2\langle m s, m' s' |V|m' s', m s \rangle \nonumber\\&&
- \langle m s, m' s' |V|m s, m' s' \rangle )
\end{eqnarray}
\subsection{$S=1$ exciton}

In each isolated dot, 
at a critical field, and driven by electron-electron  Coulomb interaction and
 increasing electron kinetic energy,
a transition from $S=0$ singlet to $S=1$ triplet is seen.
The cost in kinetic energy is lowered if localized electrons in 
one dot flip the spin ($S_L=1$), while the other electrons in the second
dot occupy the lowest energy single particle states to form spin singlet
droplet ($S_R=0$).
This configuration corresponding to $X_{RR}$ (or $X_{LL}$)
in Fig. \ref{X}, is equivalent to a localized 
electron-hole excitation in one dot in the presence of the 
background of electrons in the other dot.
Because of the geometrical symmetry associated with the electron-hole 
excitation, the ground state of the system (without any external
bias and interdot tunneling) has double degeneracy:
The state with $(S_L=0,S_R=1)$ has exactly the same energy as the
state $(S_L=1,S_R=0)$.
The many body wavefunction of such a molecular state
can be expressed as linear combination of 
degenerate states $|S_L=0,S_R=1\rangle$, and $|S_L=1,S_R=0\rangle$.
We identify these pairs of excitations as quantum Hall ferrimagnets.
For a range of magnetic fields these two states are separated
from another pair of single excitations $X_{LR}$ and $X_{RL}$ 
with $(S_L=1/2,S_R=1/2)$ by an energy gap due to  
Coulomb interaction.

An excitonic state is constructed by removing an electron
from occupied state and putting into an unoccupied state
\begin{eqnarray}
|X_{j\rightarrow i}\rangle = d^\dagger_i d_j |\nu=2\rangle
\end{eqnarray}
where $j\equiv (m,s,\downarrow)$, and $i\equiv (m',s',\uparrow)$
The lowest energy basis of the single exciton
(first spin flip state) is depicted in Fig. \ref{X}.  
Labeling the direct and indirect spin flip transitions 
(with ferrimagnetic and ferromagnetic spin ordering)
by $\{X_{RR}, X_{LL}\}$, and $\{X_{LR}, X_{RL}\}$
and using their symmetries we find
that the direct (indirect) states are two-fold degenerate  
$E_{X_{RR}} = E_{X_{LL}}$ $(E_{X_{LR}} = E_{X_{RL}})$.
Here the subscripts are defined as 
$RR\equiv(1,R,\downarrow)\rightarrow (2,R,\uparrow)$,
$LL\equiv(1,L,\downarrow)\rightarrow (2,L,\uparrow)$,
$RL\equiv(1,R,\downarrow)\rightarrow (2,L,\uparrow)$,
and
$LR\equiv(1,L,\downarrow)\rightarrow (2,R,\uparrow)$.
Note that in non-interacting system the basis is four fold degenerate 
$E_{X_{RR}} = E_{X_{LL}} = E_{X_{LR}} = E_{X_{RL}}$.

Denoting quasi-particle energy levels 
by $\varepsilon_i = \tilde{\epsilon}_i + \Sigma(i)$,
the energy of one exciton reads 
\begin{eqnarray}
E_{X_{j\rightarrow i}} = E_{\nu=2} + \varepsilon_i - \varepsilon_j 
- \langle i,j|V|j,i\rangle ,
\end{eqnarray}
where the last term is the electron-hole Coulomb interaction.
In the basis of single exciton states, the  
Hamiltonian of the QD molecules can be expressed as
\begin{eqnarray}
H_{4\times4}^{\rm eff}=T_{4\times4}^{\rm eff}+V_{4\times4}^{\rm eff},
\end{eqnarray}
where
\begin{eqnarray}
T_{4\times4}^{\rm eff}=
\left(\begin{array}{cccc}
E_{X_{RR}} & 0 & +t_1 & -t_2 \\
0  & E_{X_{LL}} & -t_2 & +t_1 \\
+t_1 & -t_2 & E_{X_{LR}} & 0 \\
-t_2 & +t_1 & 0 & E_{X_{RL}}
\end{array}\right)
\end{eqnarray}
is the non-interacting part, and 
\begin{eqnarray}
V_{4\times4}^{\rm eff}=
\left(\begin{array}{cccc}
0 & V_{RRLL} & V_{RRLR} & V_{RRRL} \\
V^*_{RRLL}  & 0 & V_{LLLR} & V_{LLRL} \\
V^*_{RRLR} & V^*_{LLLR} & 0 & V_{LRRL} \\
V^*_{RRRL} & V^*_{LLRL} & V^*_{LRRL} & 0
\end{array}\right)
\end{eqnarray}
is the Coulomb interaction matrix between the 
single exciton states.

Note that the pair of states $X_{RR}$ and $X_{LL}$ 
is not coupled by the single tunneling term
because the scattering process between $X_{RR}$ and $X_{LL}$
requires the exchange of two particles simultaneously.
For that reason this process is second order in tunneling.
The same is true for the states $X_{LR}$, and $X_{RL}$. 
\subsection{$S=2$ bi-exciton}
With increasing magnetic fields,
a higher polarized state with $S=2$,
$(S_L=1,S_R=1)$ equivalent to a bi-excitonic state
tends to appear as ground-state.
A bi-exciton is constructed by removing a pair of electrons
from occupied states and putting into unoccupied states
\begin{eqnarray}
|X_{j\rightarrow i, k\rightarrow l}\rangle 
= d^\dagger_i d^\dagger_l d_j d_k |\nu=2\rangle.
\end{eqnarray}
The energy of biexcitonic state can be decomposed into the
energy of two single excitons plus their interaction
\begin{eqnarray}
\Delta E_{XX_{j\rightarrow i, k\rightarrow l}} = 
\Delta E_{X_{j\rightarrow i}} + \Delta E_{X_{k\rightarrow l}}
+ \delta V.
\end{eqnarray}
$\delta V$ is the binding energy between two excitons,
accounted for the electron-electron, electron-hole and hole-hole
interactions
\begin{eqnarray}
\delta V &=& \langle l,i |V| i,l \rangle - \langle l,i |V| l,i \rangle \nonumber\\&&
           - \langle l,j |V| j,l \rangle                                 \nonumber\\&&
           - \langle k,i |V| i,k \rangle                                 \nonumber\\&&
           + \langle j,k |V| k,j \rangle - \langle j,k |V| j,k \rangle   
\end{eqnarray}
In the quantum dot molecule considered in this study, we find $\delta V < 0$
for large magnetic fields, i.e.,
two isolated excitons favor to pair and form a biexcitonic state where
the energy of a biexciton is lower than energy of two isolated single excitons.

\subsection{Excitonic condensation: a SP-CI effective model}
\label{ExcitonicCondensation}

The eigenvalues of $T_{4\times4}^{\rm eff}$
follow $E_{AS} < E_{SS} < E_{AA} < E_{SA}$
(as $t_2 > t_1$).
The corresponding eigenstates are
$|X_{AS}\rangle, |X_{SS}\rangle, |X_{AA}\rangle, |X_{SA}\rangle$,
shown in Fig. \ref{Xsa}.
Note that for non-interacting electrons $\Delta E_X^0 = E^0_X - E_{\nu=2} > 0$
where $E^0_X = E_{X_{RR}} = E_{X_{LL}} = E_{X_{RL}} = E_{X_{LR}}$.
By simplifying the Coulomb interaction among electrons as
$V_{ijkl} \rightarrow V_0 > 0$,
we can calculate the self-energy, and the ground state
energy of $S=0$ state analytically as: $\Sigma=NV_0/2$, and
$E_{\nu=2}=E^0_{\nu=2} + (N/2)^2 V_0$.
The latter is the energy of single exciton measured from the ground state energy.
We can also calculate the energy of single exciton
$\Delta E_{X_{j\rightarrow i}} = \epsilon_i - \epsilon_j - V_0 
< \Delta E_X^0$, and 
the energy of the biexciton 
$\Delta E_{XX} = 2\Delta E_X - 2V_0$, where
$\Delta E_X \equiv \Delta E_{X_{RR}} 
=\Delta E_{X_{LL}}$, and $\delta V = -2V_0 < 0$.
The energy difference between biexciton
and a single exciton follows $\Delta E_{XX} - \Delta E_X = 
\Delta E_X - 2V_0$.
In the limit of strong Coulomb interaction (large magnetic fields) we find
$\epsilon_i - \epsilon_j < 3 V_0$ and $E_{XX} <  E_X$.

The energies of 
two distinct  excitons
corresponding to the direct and indirect first spin flip transition, 
$X_{RR}$, and $X_{LR}$, 
the lowest energy eigenvalue of $H_{4\times 4}^{\rm eff}$,
and the energy of biexciton ($S=2$), calculated  
from the energy of the ground state with $S=0$, are shown in Fig. \ref{T1_LR}.

\begin{figure}
\begin{center}
\includegraphics[width=0.6\linewidth]{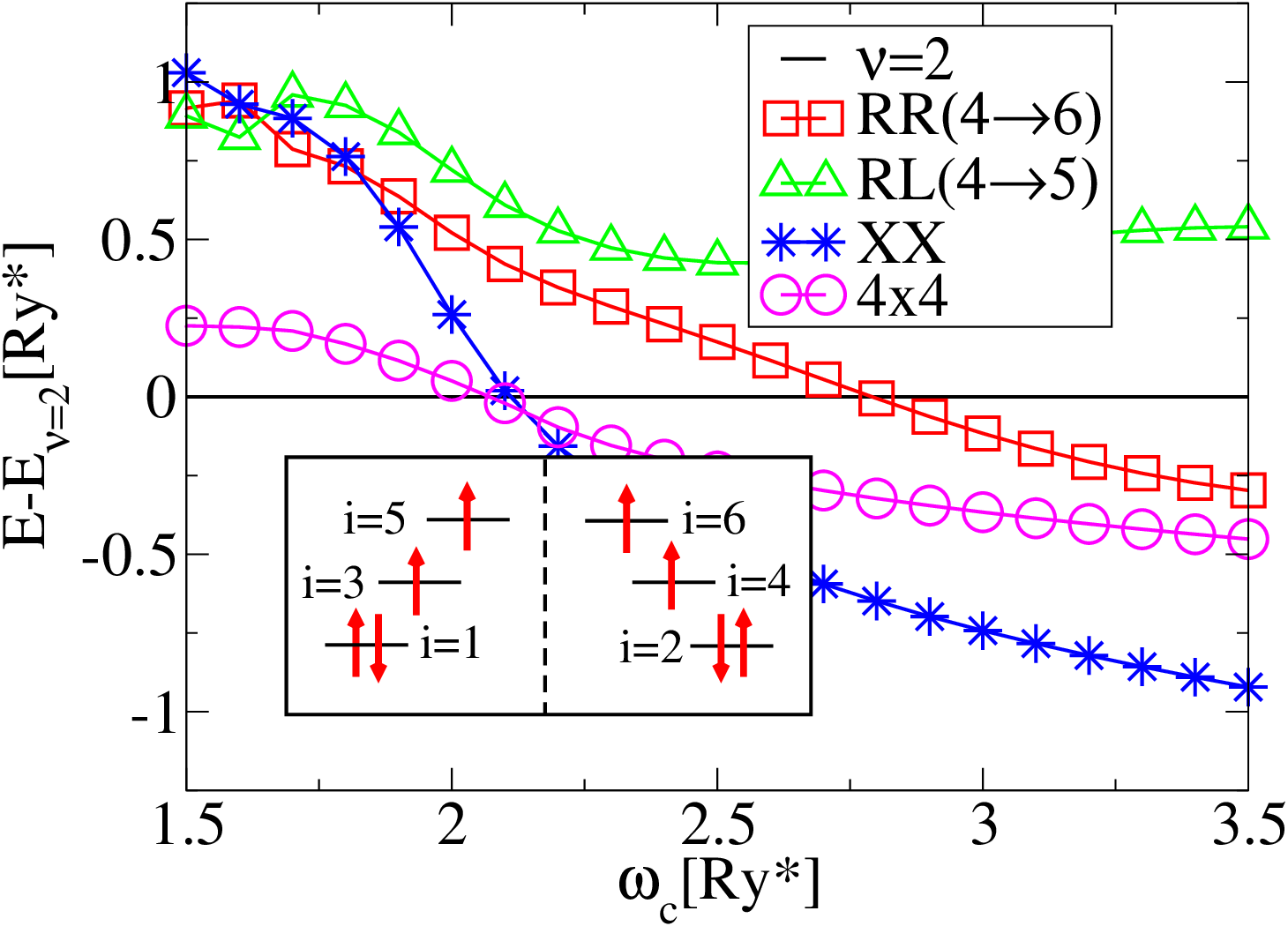}
\noindent
\caption{
The energies of spin excitons with respect to $\nu=2,S=0$ state 
as a function of magnetic field.
$E_1^{\rm eff}$ is the lowest eigen-energy of the effective Hamiltonian
$H_{4\times4}^{\rm eff}$.
The inset illustrates the energy levels of single particle localized states.
}
\label{TT1_LR}
\end{center}
\end{figure}

As predicted by single configuration SP-CI, the direct spin flip
transition takes place at $\hbar\omega^*_{c1}\approx 2.8$ where
$E_{X_{RR}} < E_{\nu=2}$, as  shown in Fig. \ref{TT1_LR}. 
Within this range of magnetic fields, the energy of a biexciton is
lower than the energy of a single exciton due to strong Coulomb
interaction. At 
$\hbar\omega^*_{c2}\approx 2.1$, 
a transition to $S=2$ state is seen due to pairing of single spin 
excitons.
However, because of strong mixing between the single excitonic states,
electron correlations improves the energy of the $S=1$ state, 
enough to bring the first spin flip transition point to 
$\hbar\omega^*_{c1}\approx \hbar\omega^*_{c2}$, where three 
states with $S=0$, $S=1$, and $S=2$ appeared to be almost degenerate.
The Zeeman coupling remove such degeneracy.
As a result, the $S=1$ state tends to become stable in a narrow 
range of magnetic field.

\section{conclusion}\label{Conclusion}
In conclusion, 
we have presented detailed analysis of the magnetic field driven spin transitions 
in quantum dot
artificial molecules with $N=8$ electrons as a function of external
magnetic field, Zeeman energy, and the detuning,
using Hartree-Fock configuration interaction method. 
The magnetic field allows the tuning of the total spin of electrons in
each artificial atom. Quantum mechanical tunneling and electron-electron interactions
couple spins of each artificial atom and result in ferromagnetic, anti-ferromagnetic , and 
ferrimagnetic states of quantum dot
artificial molecules, tunable by the magnetic field and barrier potential.

\section{acknowledgement}
Authors acknowledge the support by the NRC High 
Performance Computing project and by the Canadian 
Institute for Advanced Research, and discussions with M.Korkusinski, M. Pioro-Ladriere and A. Sachrajda.




\begin{thebibliography}{99}

\bibitem{kouwenhoven} 
W. G. van der Wiel, S. De Franceschi, J. M. Elzerman,
T. Fujisawa, S. Tarucha, L. P. Kouwenhoven,
Rev. Mod. Phys. \textbf{75}, 1 (2003).

\bibitem{Ciorga} 
M. Ciorga, A. Wensauer, M. Pioro-Ladriere, M. Korkusinski, 
J. Kyriakidis, A. S. Sachrajda, and P. Hawrylak,
Phys. Rev. Lett. {\bf 88}, 256804 (2002).

\bibitem{MichelPRL93}
M. Pioro-Ladriere, M. Ciorga, J. Lapointe, P. Zawadzki, M. Korkusinski, 
P. Hawrylak, and A. S. Sachrajda, \prl {\bf 91}, 026803 (2003).

\bibitem{petta}
J. R. Petta,  A. C. Johnson, C. M. Marcus, M. P. Hanson, and A. C. Gossard,
Phys. Rev.  Lett. {\bf 93}, 186802 (2004).

\bibitem{Korkusinski} 
M. Korkusinski, P. Hawrylak, M. Ciorga, M. Pioro-Ladriere, 
and A. S. Sachrajda, \prl {\bf 93}, 206806 (2004).

\bibitem{petta2}
J. R. Petta, A. C. Johnson, J. M. Taylor, E. A. Laird,
A. Yacoby, M. D. Lukin, C. M. Marcus, M. P. Hanson, A. C. Gossard,
Science, {\bf 309},  2180 (2005).

\bibitem{Koppens}
F. H. L. Koppens, C. Buizert, K. J. Tielrooij, I. T. Vink, K. C. Nowack, 
T. Meunier, L. P. Kouwenhoven and L. M. K. Vandersypen,
Nature, {\bf 442},  766 (2006).

\bibitem{GuyPhysicaB}
S. Sasaki, D. G. Austing, S. Tarucha, Physica B {\bf 256}, 157 (1998).

\bibitem{Kouwenhoven2} 
L. P. Kouwenhoven, D. G. Austing, and S. Tarucha, 
Rep. Prog. Phys. {\bf 64}, 701 (2001).

\bibitem{Hanson03} 
R. Hanson, B. Witkamp, L. M. K. Vandersypen, L. H. Willems van Beveren, 
J. M. Elzerman, and L. P. Kouwenhoven, 

Phys. Rev. Lett. {\bf 91}, 196802 (2003).

\bibitem{GuyPRB} 
D. G. Austing, S. Tarucha, H. Tamura, K. Muraki, F. Ancilotto, 
M. Barranco, A. Emperador, R. Mayol, and M. Pi,
Phys. Rev. B {\bf 70}, 045324 (2004).



\bibitem{JJP} 
J. J. Palacios and P. Hawrylak, Phys. Rev. B {\bf 51}, 1769 (1995).


\bibitem{qinfo}
A. Galindo and M. A. Martin-Delgado,
Rev. Mod. Phys. {\bf 74}, 347 (2002).

\bibitem{brum}
J. A. Brum, and P. Hawrylak, Superlattices Microstruct. {\bf 22}, 431 (1997).

\bibitem{Loss-DiVincenzo}
D. Loss and D.P. DiVincenzo, Phys. Rev. A {\bf 57}, 120 (1998);
G. Burkard, D. Loss and D. P. DiVincenzo,  
Phys. Rev. B \textbf{59}, 2070 (1999).

\bibitem{HuDasSarma}
Xuedong Hu and S. Das Sarma, Phys. Rev. A 64, 042312 (2001),
X. Hu and S. Das Sarma, {\em ibid} 61, 062301 (2000).

\bibitem{Kolehmainen}
J. Kolehmainen, S.M. Reimann, M. Koskinen, M. Manninen,
Eur. Phys. J. B  {\bf 13}, 731 (2000).

\bibitem{RaminWojtekPawel}
Ramin M. Abolfath, W. Dybalski, Pawel Hawrylak, 
Phys. Rev. B {\bf 73}, 075314 (2006).

\bibitem{Canted} 
L. Martin-Moreno, L. Brey, and C. Tejedor,
Phys. Rev. B {\bf 62}, R10633 (2000);
David Sanchez, L. Brey, and Gloria Platero,
Phys. Rev. B {\bf 64}, 235304 (2001).

\bibitem{SteveAllan}
S. M. Girvin and A. H. MacDonald, in {\em Perspectives in Quantum Hall Effects: 
Novel Quantum Liquids in Low-Dimensional Semiconductor Structures}, 
edited by S. Das Sarma and A. Pinczuk (Wiley, New York, 1997).

\bibitem{QHFerriPRL}
Ramin M. Abolfath, and Pawel Hawrylak, \prl {\bf 97}, 186802 (2006).

\bibitem{Sakurai}
J. J. Sakurai, {\em Modern Quantum Mechanics}, (Addison-Wesley, 2nd Ed., 1994).

\bibitem{SC grains}
Y. Nakamura, C. D. Chen, and J. S. Tsai, \prl {\bf 79}, 2328 (1997);
Y. Nakamura, Yu. A. Pashkin, and J. S. Tsai, Nature {\bf 398}, 786 (1999).

\bibitem{Ferri1} 
James Smart, {\em Effective field theories of magnetism}, 
(Saunders, Philadelphia, 1966).

\bibitem{Ferri2} 
S. Yamamoto, Phys. Rev. B {\bf 69}, 064426 (2004);
S. Yamamoto, T. Fukui, K. Maisinger, and U. Schollwöck, J. Phys.: 
Condens. Matter {\bf 10}, 11033 (1998);
M. Abolfath, H. Hamidian, and A. Langari, cond-mat/9901063 
(unpublished), and references therein.

\bibitem{Wensauer} 
Andreas Wensauer, Marek Korkusinski, and Pawel Hawrylak,
Phys. Rev. B {\bf 67}, 035325 (2003).

\bibitem{Scarola}
V. W. Scarola and S. Das Sarma, \pra {\bf 71}, 032340 (2005).

\bibitem{Szabo_book}
A. Szabo and N. S. Ostlund, 
{\em Modern~ Quantum~ Chemistry} (McGraw-Hill, New York, 1989).

\bibitem{YannouleasPRB}
Constantine Yannouleas and Uzi Landman, \prb {\bf 68}, 035325 (2003);
Constantine Yannouleas and Uzi Landman, \prb {\bf 68}, 035325 (2003);
C. Yannouleas and U. Landman, J. Phys.: Condens. Matter {\bf 14}, L591 (2002);
C. Yannouleas and U. Landman, Int. J. Quantum Chem. {\bf 90}, 699 (2002).

\bibitem{MichRamin}
M. Pioro-Ladriere, M. R. Abolfath, P. Zawadzki, J. Lapointe, 
S. A. Studenikin, A. S. Sachrajda, and P. Hawrylak,
Phys. Rev. B {\bf 72}, 125307 (2005).

\bibitem{JCP06}
Ramin M. Abolfath, and Pawel Hawrylak, 
J. Chem. Phys. {\bf 125}, 034707 (2006).

\bibitem{McCombe} 
B. D. McCombe and R. J. Wagner, \prb {\bf 4}, 1285 (1971).

\bibitem{IgorRMP} 
I. \v{Z}uti\'c, J. Fabian, and S. Das Sarma, Rev. Mod. Phys. {\bf 76}, 
323 (2004).

\end{thebibliography}
\end{document}